\documentclass{article}
\usepackage{graphicx}
\usepackage{apalike}
\usepackage{setspace}
\usepackage{jfrExamplee}

\usepackage{fancyhdr}
\usepackage{times}

\usepackage{multicol}
\usepackage{multirow}

\usepackage[bookmarks=true]{hyperref}

\usepackage{epsfig} 
\usepackage{amsmath} 
\usepackage{amssymb}  
\DeclareMathOperator{\diag}{diag}

\usepackage{algorithm,algorithmic}
\usepackage{graphicx}
\usepackage{epstopdf}
\usepackage{booktabs} 
\usepackage{array} 
\newcolumntype{P}[1]{>{\centering\arraybackslash}p{#1}}
\newcolumntype{M}[1]{>{\centering\arraybackslash}m{#1}}
\newcolumntype{L}[1]{>{\raggedright\let\newline\\\arraybackslash\hspace{0pt}}m{#1}}
\newcolumntype{C}[1]{>{\centering\let\newline\\\arraybackslash\hspace{0pt}}m{#1}}
\newcolumntype{R}[1]{>{\raggedleft\let\newline\\\arraybackslash\hspace{0pt}}m{#1}}
\usepackage[T1]{fontenc}
\usepackage[svgnames,table]{xcolor}

\usepackage{verbatim} 
\usepackage{footnote} 

\newcommand{\eg}{\textit{e.g.}}

\usepackage{pdfpages}
\usepackage{etoolbox}
\makeatletter
\patchcmd\@combinedblfloats{\box\@outputbox}{\unvbox\@outputbox}{}{%
	\errmessage{\noexpand\@combinedblfloats could not be patched}%
}%
\makeatother

\title{TrackerBots: Autonomous UAV  for Real-Time Localization and Tracking of Multiple Radio-Tagged Animals}

\author{Hoa Van Nguyen \\
School of Computer Science \\
The University of Adelaide \\
SA 5005, Australia \\
\texttt{hoavan.nguyen@adelaide.edu.au} \\
\And 
Michael Chesser \\
School of Computer Science \\
The University of Adelaide \\
SA 5005, Australia \\
\texttt{michael.chess@adelaide.edu.au} \\
\And
Lian Pin Koh \\
School of Ecology and Environmental Science \\
The University of Adelaide \\
SA 5005, Australia \\
\texttt{lianpin.koh@adelaide.edu.au} \\
\And 
S. Hamid Rezatofighi \\
School of Computer Science \\
The University of Adelaide \\
SA 5005, Australia \\
\texttt{hamid.rezatofighi@adelaide.edu.au} \\
\And  
Damith C. Ranasinghe \\
School of Computer Science \\
The University of Adelaide \\
SA 5005, Australia \\
\texttt{damith.ranasinghe@adelaide.edu.au} \\
}

\begin{document}
\includepdf[pages={1}]{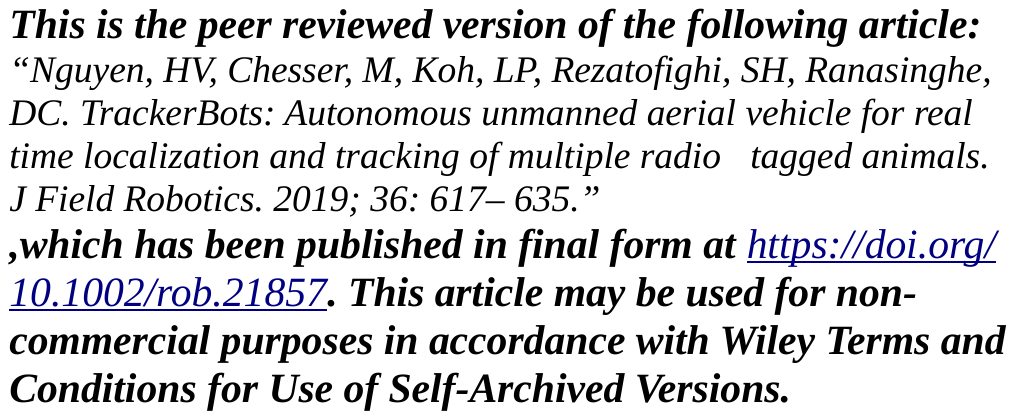}
	\maketitle
	
\begin{abstract}
	Autonomous aerial robots provide new possibilities to study the habitats and behaviors of endangered species through the efficient gathering of location information at temporal and spatial granularity not possible with traditional manual survey methods. We present a novel autonomous aerial vehicle system---\textit{TrackerBots}---to track and localize multiple radio-tagged animals.
	The simplicity of measuring the received signal strength indicator (RSSI) values of very high frequency (VHF) radio-collars commonly used in the field is exploited to realize a low cost and lightweight tracking platform suitable for integration with unmanned aerial vehicles (UAVs).
	Due to uncertainty and the non-linearity of the system based on RSSI measurements, our tracking and planning approaches integrate a particle filter for tracking and localizing and a partially observable Markov decision process (POMDP) for dynamic path planning. This approach allows autonomous navigation of a UAV in a direction of maximum information gain to locate multiple mobile animals and reduce exploration time; and, consequently, conserve on-board battery power. We also employ the concept of a search termination criteria to maximize the number of located animals within power constraints of the aerial system.	We validated our real-time and online approach through both extensive simulations and field experiments with five VHF radio-tags on a grassland plain.  
\end{abstract}

\section{Introduction}

Understanding basic questions of ecology such as how animals use their habitat, their movements and activities are necessary for addressing numerous environmental challenges ranging from invasive species to diseases spread by animals and saving endangered species from extinction. Conservation biologists, ecologists as well as natural resource management agencies around the world rely on numerous methods to monitor animals. 
Traditional methods using radio-tagging species of interest~\cite{cochran1963radio,kenward2000manual} 
as well as more recent vision-based sensors~\cite{selby2011autonomous,olivares2015towards} or infrared (IR) based sensors~\cite{zhou2013thermal,christiansen2014automated,gonzalez2016unmanned,ward2016autonomous} are employed for these tasks. 
IR-based sensors are sensitive to environmental temperature and become less reliable when they are used outdoors, especially during day-times in summer months~\cite{zhou2013thermal}.  In general, vision-based approaches are less effective when animals are camouflaged and are susceptible to visual occlusions, e.g. by grass, shrubs and even other animals. Most significantly, due to the difficulty of automatically recognizing individual animals using vision/IR based approaches, tracking multiple animals with these sensors require dealing with the very challenging problem of data association~\cite{bar1987tracking,stone2013bayesian}. Often, conservation biologists need tools to track and monitor a specific set of individual animal species; for example, individuals of a reintroduction species into a natural habitat. This becomes difficult to achieve in the presence of occlusions and data associations problems of IR/vision based approaches. 
Thus, capturing and collaring concerned species with Very High Frequency (VHF) radio tags and the subsequent use of VHF telemetry or radio tracking is the most important and cost-effective tool employed to study the movement of a wide range of animal sizes~\cite{wikelski2007going} in their natural environments~\cite{kays2011tracking,thomas2012wildlife,tremblay2017low,webber2017radio}. 

However, the traditional method of radio tracking is not without its problems. Tracking radio-collared animals typically requires researchers to trek long distances in the field, armed with cumbersome VHF radio receivers with hand-held antennas and battery packs to manually home in on radio signals emitted from radio-tagged or collared animals. Consequently, the precious spatial data acquired through radio tracking come at a significant cost to researchers in terms of manpower, time and funding. The problem is often compounded by other challenges, such as low animal recapture rates, equipment failures, and the inability to track animals that move into inaccessible terrains. Furthermore, many of our most endangered species also happen to be the most difficult to track due to their small size, inconspicuousness, and location in remote habitats. 
	
\begin{figure}[!tbp]  
		\centering
		\includegraphics[width=14cm]{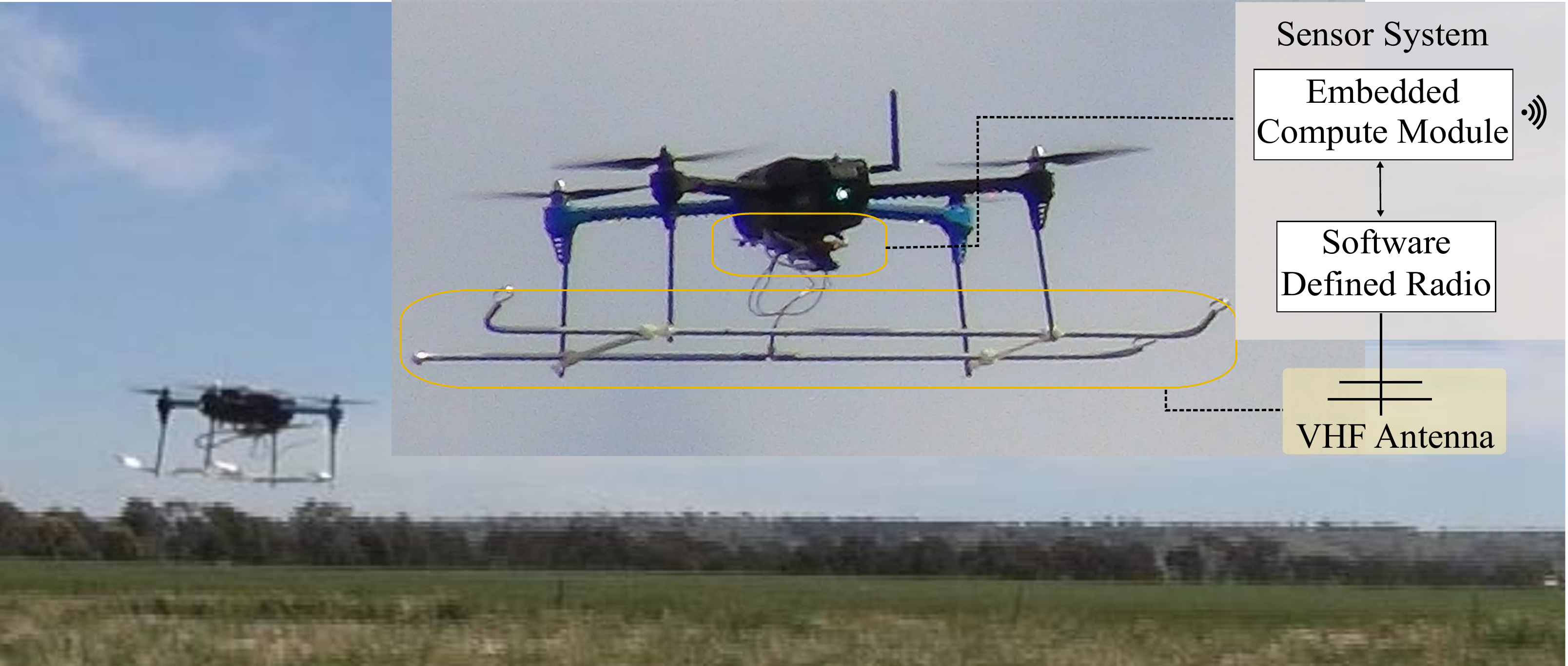}
		\caption{\textit{TrackerBots}: An overview of the UAV tracking platform with its sensor system.}
		\label{fig_UAV_System_Overview}
	\end{figure}

Automated tracking and location of wildlife with autonomous unmanned aerial vehicles (UAVs) can provide \textit{new} possibilities to better understand ecology and our native wildlife to safeguard biodiversity and manage our natural resources cost-effectively. We present a low-cost approach capable of realization in a lightweight payload for transforming existing commodity drone platforms into autonomous aerial vehicle systems as shown in Fig. \ref{fig_UAV_System_Overview} to empower conservation biologists to track and localize multiple radio-tagged animals. 
	
	The main contribution of our work is a new autonomous aerial vehicle system for simultaneously tracking and localizing multiple mobile radio-tagged animals using VHF radio-collars, commonly used in the field.  In particular: 

\begin{itemize}
\item Our system is realized in a 260~g payload suitable for a multitude of low-cost, versatile, easy to operate multi-rotor UAVs. Our lightweight realization---of less than 2~kg system mass---is achieved through a new sensor design that exploits the simplicity of a software-defined radio architecture for capturing received signal strength indicator (RSSI) value from multiple VHF radio tags and a compact, lightweight VHF antenna geometry. The lightweight design is significant for achieving longer flight times on a given UAV as well as making the technology more accessible in jurisdictions, such as Australia, where systems under 2~kg can be flown without a pilot license.
\item  We formulate a \textit{joint} tracking and path planning problem to realize a real-time and online autonomous system. Due to the noisy, complex and nonlinear characteristics of RSSI data, we integrate a sequential Monte Carlo implementation of a Bayesian filter, also known as particle filter (PF), for real-time tracking and localization \textit{jointly} with a partially observable Markov decision process (POMDP) for modeling a path planning decision process. We evaluate information based reward functions to evaluate control actions for path planning. We use R\'{e}yni divergence between prior and posterior estimates of target locations for autonomy and dynamic online path planning to minimize flight time while maximizing the number of located animals. Further, our formulation considers the trade-off between location accuracy and resource constraints of the UAV, its maneuverability, and power constraints to develop a practical solution.
\item  We validate our method through extensive simulations and field experiments with mobile VHF radio-tags. In particular, we conducted: \textit{i)} over 10 manual flights to both evaluate and measure the performance of our sensor system; and \textit{ii)}  we performed 20 autonomous flights under two different settings with a mix of target dynamics to demonstrate the robustness and scalability of our approach. To the best of our knowledge, ours is the first demonstration of an autonomous online aerial robot system for tracking and locating multiple mobile VHF radio-tags in real-time.
\item In order to support researchers in the field and facilitate the adoption of new technologies in the field, we provide a complete design description of \textit{TrackerBots}, including a repository of source code to develop our fully autonomous system~\footnote{see: \url{https://github.com/AdelaideAuto-IDLab/TrackerBots}
}
\end{itemize}

\section{Related Work}

Our problem is embedded in the development of a UAV planning method for tracking multiple mobile radio-tagged objects using the simplicity of received signal strength measurements. Therefore: \textit{i)} we review studies in the field of received signal strength measurement based tracking with a specific focus on methods developed for UAVs and wildlife tracking; \textit{ii)} we review multi-target tracking methods since our problem involves tracking multiple radio-tagged targets; and \textit{iii)} we focus on related work in the field of tracking radio-collared animals using UAVs.

\noindent\textbf{Received signal strength indicator (RSSI)-based Tracking:} This method is studied in localizing objects in both indoor and outdoor environments. The approach relies on using the strength of a radio signal from an emitter captured by a receiver to estimate, for example, the distance to the emitter.  Related methods with possible applications to wildlife tracking can be found in the use of wireless sensor networks (WSN) for tracking a radio wave emitter. In~\cite{Caballero2008,Sarkka2014} a mobile beacon is localized by a fixed number of sensor nodes with known locations. 
The first automated VHF telemetry measurement system was reported in~\cite{kays2011tracking}. A set of six ground-based antenna arrays deployed in a rainforest localized radio-tagged animal locations using bearing estimates obtained from signal strength measurements made by ground-based stations. 
These methods are advantageous for meeting long-term monitoring needs. However, the scale of the fixed and powered infrastructure required prior to a tracking task and the cost of deployment and maintenance over a large area make these approaches difficult for general use cases. In contrast, a UAV based measurement method can provide greater flexibility and a lower cost approach. 
Off-line estimations of a radio-tag's location obtained from signal strength data logged from a UAV was demonstrated in~\cite{jensen2014monte}. Developments in Software-defined radios (SDRs) have enabled new capabilities to process multiple radio-tag signals simultaneously. 
Early efforts to demonstrate the possibility of incorporating SDR architectures with a UAV to detect multiple transmitted signals from radio-tags were reported in~\cite{dos2014small,vonehr2016software}. Notably, the studies above with UAVs were performed under the assumption of stationary radio tags. The task of autonomously tracking and locating multiple mobile radio-tagged targets from a UAV remains.

\noindent\textbf{Multi-target Tracking:} The objective of multi-target tracking is to accurately estimate the unknown state of multiple objects or targets using noisy observations. The basic problem in multi-target tracking is the unknown associations between measurements and targets~\cite{bar1987tracking}. Traditional multi-target tracking formulations include multiple hypotheses tracking (MHT) filter~\cite{blackman1986multiple}, the joint probabilistic data association filter (JPDAF)~\cite{bar1987tracking}, and the probabilistic MHT (PMHT) filter~\cite{streit1994maximum}. These approaches require explicit associations between measurements and targets and propagate these hypotheses over time~\cite{vo2006gaussian}. Another alternative approach which obviates explicit data associations uses finite set statistics (FISST) proposed by Mahler based on Random Finite Set (RFS) theory. This formulation has gained considerable interest in recent years and has lead to a number of new filtering solutions such as the probability hypothesis density (PHD) filter~\cite{mahler2003phd}, the cardinalized PHD (CPHD) filter \cite{mahler2007cphd}, the multi-object multi-Bernoulli (MeM-Ber) filter \cite{mahler2007statistical,vo2009cardinality}, the labeled multi-Bernoulli (LMB) filter \cite{reuter2014lmb}, and the generalized labeled multi-Bernoulli (GLMB) filter \cite{vo2013glmb,vo2014glmb}. Since the radio-tagged methods provide an elegant solution where each target  can be uniquely identified by its transmitted signal, our problem does not suffer from the data association problems mentioned above. Thus, we propose formulating a Particle Filter (PF) ~\cite{Gordon1993} for tracking radio-tags. This is similar to the approach in ~\cite{charrow2015active}. In contrast to the simulation-based study of indoor robots in~\cite{charrow2015active}, we design and implement our algorithm on a UAV with a sensor system for obtaining RSSI-based measurements of multiple radio-tagged objects in outdoor environments.

\noindent\textbf{UAV-based Autonomous Localization and Tracking:}
Since this application is related to locating VHF collared animals, we will focus on progress made towards the autonomous localization and tracking of multiple VHF radio-tagged animals here.

Pioneering achievements in autonomous wildlife tracking have been made through simulation studies~\cite{Posch2009} and experimentally demonstrated systems ~\cite{cliff2015online,Korner2010,tokekar2010robotic,vander2014cautious} in recent years. In particular, the first demonstration of a UAV was  presented in~\cite{cliff2015online}. 
	
The recent approaches~\cite{cliff2015online,vander2014cautious} for real-time localization of a static target (assuming stationary wildlife) used wireless signal characteristics captured by a narrowband receiver to estimate location; in particular, the angle-of-arrival (AoA) of a radio beacon was determined using an array of antennas with the information related to a ground-based receiver for location estimations. Although the approach can conveniently manage topological variations in terrain, AoA systems require a large bulky receiver system and multiple antenna elements as well as long observation times; 45 seconds per observation as reported in \cite{cliff2015online}. Moreover, the antenna systems being mounted on top of the UAV~\cite{cliff2015online} is likely to lead to difficulty in tracking terrestrial animals although being suitable for locating avian species dwelling in trees.

\noindent\textbf{Summary:} We can see that there are few investigations that have studied the problem of locating radio-collared animals using autonomous robots. Although a system based on angle-of-arrival was recently evaluated to locate a stationary animal, the development of a low-cost and lightweight autonomous system capable of long-range flights and localization of multiple \textit{mobile} radio-collared animals still remains. 

We present an alternative approach exploiting RSSI based measurements because of the ability to use a simpler sensing system on board commodity UAVs to realize lower cost and longer flight time UAVs for tracking and localizing multiple animals. Together with a theoretical framework for joint tracking and planning, we design, build and demonstrate a lightweight autonomous aerial robot platform. Our robot platform has the potential to provide a cost-effective method for wildlife conservation and management. To the best of our knowledge, ours is the first demonstration of an autonomous online aerial robot system for tracking and locating multiple mobile VHF radio-tags in real-time.

\section{Tracking and Planning Problem Formulation}
Real-time tracking requires an online estimator and a dynamic planning method. 
This section presents our tracking and localizing formulation under the theoretical frameworks of a Bayesian filter for tracking and POMDP for planning strategy.

\subsection{Tracking and localizing}\label{sec_trackign_localizing}
For tracking, we use a Bayesian filter. It is an online estimation technique which deals with the problem of inferring knowledge about the unobserved state of a dynamic system---in our problem, wildlife---which changes over time, from a sequence of noisy measurements. Suppose $\mathbf{x} \in \mathcal{X}$ and $\mathbf{z} \in \mathcal{Z}$ are respectively the system (kinematic) state vector in the state space $\mathcal{X}$ and the measurement (observation) vector
in the observation space $\mathcal{Z}$. The problem is estimating the state \( \mathbf{x} \in \mathcal{X} \) from the measurement \( \mathbf{z} \in \mathcal{Z} \)  or calculating the marginal posterior distribution \(\mathit{p(\mathbf{x}_\mathit{k} | \mathbf{z}_{1:k})}\) sequentially through \textbf{\textit{prediction}} \eqref{eq:chapman} and \textbf{\textit{update}} \eqref{eq:bayesfilter} steps.

\begin{align}
	\mathit{p(\mathbf{x}_\mathit{k} | \mathbf{z}_\mathit{1:k-1})} &= \mathit{\int{p(\mathbf{x}_\mathit{k} | \mathbf{x}_\mathit{k-1}) p(\mathbf{x}_\mathit{k-1} | \mathbf{z}_\mathit{1:k-1}) d\mathbf{x}_\mathit{k-1 } }}, \label{eq:chapman} \\
	\mathit{p(\mathbf{x}_\mathit{k} | \mathbf{z}_{1:k})}  &= \mathit{\dfrac{p(\mathbf{z}_\mathit{k} | \mathbf{x}_\mathit{k})p(\mathbf{x}_\mathit{k} | \mathbf{z}_\mathit{1:k-1})}{\int{p(\mathbf{z}_\mathit{k} |\mathbf{x}_\mathit{k})p(\mathbf{x}_\mathit{k}|\mathbf{z}_\mathit{1:k-1})d\mathbf{x}_k}}}.  \label{eq:bayesfilter}
\end{align}

In the case of a nonlinear system or non-Gaussian noise, there is no general closed-form solution for the Bayesian recursion and $\mathit{p(\mathbf{x}_\mathit{k} | \mathbf{z}_{1:k})}$ generally has a non-parametric form. Therefore, in our problem, we use a particle filter implementation as an approximate solution for the Bayesian filtering problem due to our highly nonlinear measurement model.

\textbf{Particle Filter (PF):} \label{sec_PF} 
A particle filter uses a sampling approach to represent the non-parametric
form of the posterior density $\mathit{p(\mathbf{x}_\mathit{k} | \mathbf{z}_{1:k})}$. The samples from the distribution
are represented by a set of particles; each particle has a weight assigned to
represent the probability of that particle being sampled from the probability density function. Then, these particles representing the non-parametric form of $\mathit{p(\mathbf{x}_\mathit{k} | \mathbf{z}_{1:k})}$
are propagated over time. 
In the simplest version of the particle filter, known as the bootstrap filter first introduced by Gordon in~\cite{Gordon1993}, the samples are directly generated from the transitional dynamic model. Then, to reduce the particle degeneracy, resampling and injection techniques are implemented; a detailed algorithm can be found in~\cite{Ristic2004}. 

\textbf{Measurement model: }
The update process of a PF requires the derivation of a likelihood of measurements. In our problem, based on estimating a target's---VHF radio tag's---range from the receiver, we require a realistic signal propagation model to obtain the likelihood of receiving a given measurement. We employ two VHF signal propagation models suitable for describing RSSI measurements in non-urban outdoor environments \mbox{\cite{patwari2005locating,wc1974microwave}}.
Denoting $\mathbf{h(x,u)}$ as the RSSI measurement function between target $\mathbf{x}$ and observer (UAV) state $\mathbf{u}$, we have:

\noindent\textbf{\textit{i)}} \textbf{\textit{Log Distance Path Loss Model (LogPath)}:} The received power is the only line of sight power component transmitted from a transmitter subjected to signal attenuation such as through absorption and propagation loss \cite{patwari2005locating}: 

\begin{equation} \label{eq_LogPath}
	\begin{aligned}
		\mathbf{h(x,u)} = P_r^{d_0}- 10n\log_{10}(d(\mathbf{x},\mathbf{u}_p)/d_0) + G_r(\mathbf{x,u}), 
	\end{aligned}
\end{equation}
\noindent where 
\begin{itemize}	
	\item $\mathbf{x} = [p_x^{t}, p_y^{t}, p_z^{t}]^{T}$ is the target's position; $\mathbf{u}_p = [p_x^{u}, p_y^{u}, p_z^{u}]^T$ is the observer's (UAV) position in Cartesian coordinates; $\mathbf{u} = [\mathbf{u}_p;  \theta^{u}]$ is the UAV's state which includes its heading angle $\theta^{u}$. 
	\item $d(\mathbf{x},\mathbf{u}_p)$ is the Euclidean distance between the target's position and UAV's position.
	
	\item $G_r(\mathbf{x,u})$ is the UAV receiver antenna gain which depends on its heading, its position, and target's position (details explained in Sec. \ref{sec_emperical_meas}). 
	\item $P_r^{d_0}$ is received power at a reference distance $d_0$.
	\item $n$ is the path-loss exponent that characterizes the signal losses such as absorption and propagation losses and this parameter depends on the environment with typical values ranging from 2 to 4 \cite{patwari2005locating}. 
\end{itemize}
\noindent\textbf{\textit{ii)}}\textbf{\textit{ Log Distance Path Loss Model with Multi-Path Fading (MultiPath)}:} The received power is composed of both line of sight power component transmitted from a transmitter and the multi-path power component reflected from the ground plane subjected to signal attenuation such as through absorption and propagation loss: \cite[p.~81]{wc1974microwave}:

\begin{equation} \label{eq_MultiPath}
	\begin{aligned}
		\mathbf{h(x,u)} &= P_r^{d_0}- 10n\log_{10}(d(\mathbf{x},\mathbf{u}_p)/d_0) \\
		&+ G_r(\mathbf{x,u}) + 10n\log_{10}(|1+\Gamma(\psi) e^{-j\triangle \varphi}|),
	\end{aligned}
\end{equation}

\noindent where, in addition to terms in~\ref{eq_LogPath} 
\begin{itemize}
	\item $\psi$ is the angle of incidence between the reflected path and the ground plane.
	\item $\Gamma(\psi) = [\sin(\psi) - \sqrt{\varepsilon_g - \cos^2(\psi)}]/[\sin(\psi) + \sqrt{\varepsilon_g - \cos^2(\psi)}]$ is the ground reflection coefficient with $\varepsilon_g$ is the relative permittivity of the ground.
	\item $\triangle \varphi = 2\pi \triangle d / \lambda$ is the phase difference between two waves where $\lambda$ is the wavelength and $\triangle d = ((p_x^{t} - p_x^{u})^2 + (p_y^{t} - p_y^{u})^2 + (p_z^{t} + p_z^{u})^2)^{1/2} -d(\mathbf{x}, \mathbf{u}). $
\end{itemize}

In non-urban environments, received power is usually corrupted by environmental noise, with the assumption that the noise is white, the total received power $\mathbf{z} = P_r(\mathbf{x,u})$~[dBm] is:
\begin{align} \label{eq_pathloss}
	\mathbf{z}  &= \mathbf{h(x,u)} + \eta_{P}, 	
\end{align}
where $\eta_{P} \sim \mathcal{N}(0,\sigma_P^2)$ is Gaussian white noise with covariance $\sigma_P^2$. Notably, even if RSSI noise is not completely characterized by a white noise model, we found it practical to characterize the received noise with a white Gaussian model as shown in Fig. \ref{fig_Meas_Model}. 

We use data captured in experiments using our sensor system to validate the physical sensor characteristics $G_r(\mathbf{x,u})$ (see Sec. \ref{sec_Antenna_Gain}) and $n$ defined by environmental characteristics, as well as estimate the propagation model reference power parameter  $ P_r^{d0}$ and noise $\sigma_P$ (see Sec. ~\ref{sec_emperical_meas}).

\textbf{\textit{Measurement likelihood}:} Based on~\eqref{eq_pathloss} with Gaussian noise $\eta_P$, the likelihood of measurement $\mathbf{z}_k$, given target and sensor position are $\mathbf{x}_{k}$ and $\mathbf{u}_{k}$, respectively, at time $k$ is
\begin{align}
	p(\mathbf{z}_{k}|\mathbf{x}_k, \mathbf{u}_{k}) \sim \mathcal{N}(\mathbf{z}_{k};\mathbf{h}(\mathbf{x}_k,\mathbf{u}_k) , \sigma^2_P ),	
\end{align}
where $\mathcal{N}(\cdot;\mu,\textcolor{black}{\sigma^2} )$ is a normal distribution with  mean  $\mu$  and \textcolor{black}{covariance $\sigma^2$}. 

\subsection{Path Planning}
The UAV planning problem is similar to the problem of an agent computing optimal actions under a partially observable Markov decision process (POMDP) to maximize its reward. 
\cite{KAELBLING199899} have shown that a POMDP framework implements an efficient and \textit{optimal} approach based on previous actions and observations to determine the true world states. A POMDP in conjunction with a particle filter provides a principled approach for evaluating planning decision to realize an autonomous system for tracking. 

In general, a POMDP is described by the 6-tuple \( (\mathcal{S,A,T,R,O,Z}) \) where \(\mathcal{S} \) is a set of both UAV and target states ($\mathbf{s} = \{\mathbf{x,u}\} \in S$), \(\mathcal{A} \) is a set of UAV actions, \( \mathcal{T} \) is a state-transition function \(\mathcal{T}(\mathbf{s,a,s'}) = p(\mathbf{s'|s,a})\) for a given action $\mathbf{a}$, \(\mathcal{R}(\mathbf{a})\) is a reward function, \(\mathcal{O} \) is a set of observations and \(\mathcal{Z} \) is an observation likelihood 
\(\mathcal{Z}(\mathbf{s,a,o}) = p(\mathbf{o|s,a})\) with \(\mathbf{s,s'} \in \mathcal{S}\) is the current state and next state respectively, \(\mathbf{a} \in \mathcal{A} \) is the taken action and \(\mathbf{o} \in \mathcal{O} \)  is the observation---i.e., measurement. The goal of a POMDP is to find an optimal policy to maximize the total expected reward \textcolor{black}{\(\mathbb{E}[\sum_{\kappa=k+1}^{k+H} \gamma^{\kappa-k-1}\mathcal{R}_{\kappa}(\mathbf{a}_k)]\)} where \(H\) is look-ahead horizon steps, \(\gamma \) is the discount factor which serves as the value difference between the current reward versus the future reward; \textcolor{black}{\(\mathbf{a}_{k}\) is action at time step \(k\)} and $\mathbb{E}[\cdot]$ is the expectation operator \cite{Hsu2008}.

The reward function can be calculated using different methods such as task-driven or information-driven strategies. When uncertainty is high, the information gain approach is preferable to reduce a target's location uncertainty \cite{beard2017void}; hence, we used this method to calculate our reward function. There are several approaches to evaluate information gain in robotic path planning such as Shannon entropy~\cite{cliff2015online}, Kullback-Leibler (KL) divergence or R\'{e}nyi divergence~\cite{Hero2008}. We adopted the approach in~\cite{Ristic2013,Ristic2010} to implement R\'{e}nyi divergence as our reward function since it fits naturally with our Monte-Carlo sampling method. Here, R\'{e}nyi divergence measures the information gain between prior and posterior densities \cite{beard2017void,Ristic2010}:

\begin{equation} \label{ReyniDivergence}
	\begin{aligned}
		\mathcal{R}^{(i)}_{k+H}(\mathbf{a}_k) &= \dfrac{1}{\alpha-1} \log \int
		\begin{bmatrix}
			p(\mathbf{x}_{k+H}|\mathbf{z}_{1:k})
		\end{bmatrix}^{\alpha} 
		\begin{bmatrix}
			p(\mathbf{x}_{k+H}|\mathbf{z}_{1:k},\mathbf{z}^{(i)}_{k+1:k+H}(\mathbf{a}_k))
		\end{bmatrix}^{1-\alpha}
		d\mathbf{x}_{k+H},
	\end{aligned}
\end{equation}
where $\alpha \geq 0$ is a scale factor to decide the effect from the tails of two distributions. The prior density  $p(\mathbf{x}_{k+H}|\mathbf{z}_{1:k})$ is calculated by propagating current posterior particles sampled from $p(\mathbf{x}_{k}|\mathbf{z}_{1:k})$ to time $k+H$ using the prediction step~\eqref{eq:chapman}. The posterior density $p(\mathbf{x}_{k+H}|\mathbf{z}_{1:k},\mathbf{z}^{(m)}_{k+1:k+H}(\mathbf{a}_k))$ where $\mathbf{z}^{(m)}_{k+1:k+H}(\mathbf{a}_k)$ is the \textit{future} measurement set that will be observed if action $\mathbf{a}_k \in \mathcal{A}_{k}$ is taken; this is calculated by applying both prediction~\eqref{eq:chapman} and update steps~\eqref{eq:bayesfilter} up to time $k+H$. However, using Bayes update procedure is computationally expensive and prohibitive in a real-time setting. Instead, we implement a computationally efficient approach using a black box simulator proposed in~\cite{silver2010monte} along with the Monte Carlo sampling approach. Hence, the problem transforms to find an optimal action \(\mathbf{a}^*_k\in \mathcal{A}_{k} \) to maximize total expected reward:
\begin{align}
	\mathbf{a}^*_k \approx \arg \max\limits_{\mathbf{a}_k \in \mathcal{A}_{k}} \dfrac{1}{M_s} \sum_{t=k+1}^{k+H}  \sum_{m=1}^{M_s} \gamma^{t-k}\mathcal{R}^{(m)}_{t}(\mathbf{a}_k),
\end{align}
where \(M_s\) is the number of future measurements. 

\subsection{Multi-targets Tracking}
The particle filter proposed in Sec. \ref{sec_PF} can be extended to multi-target tracking (MTT). However, MTT normally deals with the complex data association problem where it is difficult to determine which measurement belongs to which target. In contrast, for our system, each target can be estimated from the measurement based on the signal frequency and tracked independently. Thus, we do not need to solve the data association problem. 
Notably, not all of the targets are detected due to, for example, the UAV movements, the measurement range limits imposed by propagation losses and receiver sensitivity. Therefore, if the target is not detected, the solver does not update its estimated position. 

Besides maximizing the number of targets localized and tracked, we formulated a termination condition for each target to conserve UAV battery power; a target is considered localized if its location uncertainty, determined by a determinant of its particles covariance, is sufficiently small (<\(N_{Th}\)). Then, those found targets are \textit{forgotten} to aid the solver to prioritize its computing resources on those targets with high uncertainty.

\section{System Implementation}
\begin{figure}[!tb] 
	\centering 
	
	\includegraphics[clip, width=17cm]{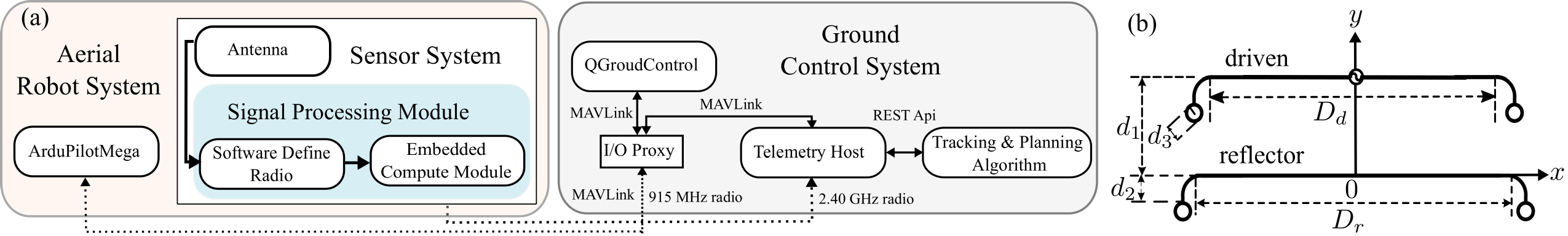}
	
	\caption{a) The communication channels between UAV and the ground control system with its main software components and protocols. The solid lines represent the internal connections/communications within the Sensor System and the Ground Control System. The dotted lines are connections between wireless interfaces such as the Aerial Robot System and the Ground Control System through two different radio channels: $915$~MHz and $2.4$~GHz. 
		b) The folded 2-element Yagi antenna design used in our sensor system for observations.} 
	
	\label{fig_Combine_PC_UAV_Communication_And_Antenna_Design}
\end{figure}

We implemented an experimental aerial robot system based on our tracking and planning formulation. An overview of the complete system is described in Fig. \ref{fig_Combine_PC_UAV_Communication_And_Antenna_Design}a. Our experimental system used a 3DR IRIS+ UAV platform and a new sensor system built with: \textit{i)} a compact directional VHF antenna design, and \textit{ii)} a software-defined signal processing module capable of simultaneously processing signals from multiple targets and remotely communicating with a ground control system for tracking and planning.
In our system,  the ArduPilotMega (APM) on the IRIS+ UAV transmits back its global positioning system (GPS) location to the Telemetry Host tool developed by our group to communicate with the APM  module using the MAVLink protocol over a 915 MHz full duplex radio channel. The sensor system together with the antenna, SDR receiver, and the embedded compute module delivers targets' RSSI data through a 2.40~GHz radio channel to the ground control system.

GPS locations of the UAV platform and targets' RSSI data are delivered to our tracking and planning algorithm---\textit{solver}---through the Telemetry Host using a RESTful web service. The solver estimates target locations and calculates new control actions per each POMDP cycle to command the UAV through MAVLink to fly to a new location. In order to ensure safety and meet University regulatory requirements, we also employ QGroundControl---a popular cross-platform flight control and mission planning software---to monitor and abort autonomous navigation. We detail our Sensor System below. 

\begin{figure}[tb] 
	\centering 
	\includegraphics[clip, width=17cm]{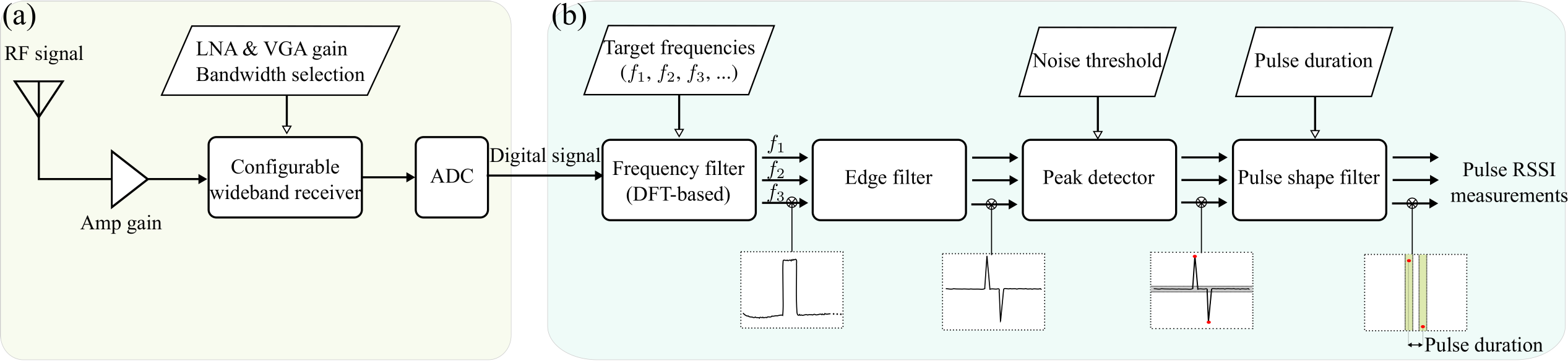}
	\caption{The signal processing module. (a) \textit{Software-defined radio:} raw input RF signals are processed through the HackRF One SDR device with different configurable amplifiers--Low Noise Amplifier (LNA) and Variable Gain Amplifier (VGA), and an ADC to convert analogue signals to digital signals. (b) \textit{Embedded compute module:} digital signals are processed on an Intel Edison board using a DFT (Discrete Fourier Transform)-based frequency filter with configurable input frequencies, edge filter and peak detector algorithms to derive radio collar RSSI measurements. } 
	\label{fig_dsp_chains}
\end{figure}

\textbf{Signal Processing Module:} Fig. \ref{fig_dsp_chains} illustrates the  components of the proposed signal processing module. 
We propose using a \textit{software-defined radio} (SDR) receiver to implement the signal processing components. The key advantages of our choice are the ability to: \textit{i)} reduce the weight of the receiver; \textit{ii)} rapidly scan a large frequency spectrum to track multiple animals beaconing on different VHF frequency channels;  and \textit{iii)} reconfigure the system on the fly because the signal processing chain is defined in software. 

In this work, we use the \textit{HackRF One} SDR---an open source platform developed by \cite{Ossmann2015} capable of directly converting radio frequency (RF) signals to digital signals using an analog-to-digital converter (ADC)---together with an Intel Edison board as our embedded compute module. We implemented a Discrete Fourier Transform (DFT) filter to isolate, from multiple signals, each unique VHF frequency channel associated with an animal radio collar and measure the signal strength of the received signal. 

\textbf{Antenna:} \label{sec_Antenna_Gain} A lightweight folded 2-element Yagi antenna was specially designed for our sensor system. Our design achieves a low profile antenna capable of being within the form factor of low-cost commodity UAVs suitable for easy operation in the field. Similar to a standard 2-element Yagi antenna, the folded design has one reflector and one driven element as shown in Fig. \ref{fig_Combine_PC_UAV_Communication_And_Antenna_Design}b. 

The antenna operates in the frequency range from $145$~MHz to $155$~MHz (a typical range for wildlife radio tags), and a center frequency of $f=150$~MHz. The length of driven and reflector elements are  $D_d= 0.3975 \lambda$ and $\ D_r = 0.402\lambda$, respectively,  
while $d_1=0.1\lambda$, $d_2 = 0.03\lambda$ and the inductive loading ring diameter is $d_3 = 0.015\lambda$. Here, the wavelength $\lambda = c/f = 2$~m with $c=3\cdot10^8$~m/s. The antenna gain model calculated for the the design is shown in Fig. \ref{Combine_RotorNoise_AntennaGain_MeasureModel}b.  
\subsection{Planning implementation for a real-time system}

Implementing planning algorithms on any real-time systems is always challenging because of its high computational demand.   Thus, in the following, we present the approaches to minimize the planning computational time while not sacrificing the overall localization performance:
\begin{enumerate}
\item Notably, for RSSI data, the uncertainty in the estimation of a target's location is reduced when the maximum gain of the directional antenna mounted on the UAV points toward the target position. Hence, to increase the localization accuracy, the UAV heading angle $\theta^{u}_{k}$ must be controlled during the path planning process, although the multi-rotor UAV can be maneuvered without changing its heading. We select a set of discrete UAV rotation angles for the control actions \(\mathcal{A}_{k} \) based on a simulation-based study to reduce the computational complexity of the POMDP planning process by limiting the number of possible actions to evaluate.
	
\item The solver performs planning in every \(N_{p}\) observation cycles with \(N_{p} > 1\) instead of every observation. This approach helps to ensure that the solver prioritizes its limited computational resource on tracking targets instead of only performing planning steps.
	
\item A coarse planning interval $t_p$ in the planning procedure is implemented to minimize the computational time by reducing the number of look-ahead steps while still having the same look-ahead horizon. For example, if
	we want to estimate the target's state in a 10 second horizon, we can use the normal interval $t_{p}= 1$~s and estimate the target's state 10 times or use the coarser interval $t_{p} = 5$~s and perform the estimation twice; the latter approach is computationally less expensive.
	
\item Instead of selecting the best action from the possible action space $\mathcal{A}_k$, the domain knowledge of the receiver antenna gain is used to select a subset of actions that give the highest received gain using \textbf{Alg.}~\ref{Algo_UAV_SubSet_Possible_Location}.
\end{enumerate}	
	
\begin{algorithm} [!htbp] 
	\caption{Calculate the control action subset}
	\label{Algo_UAV_SubSet_Possible_Location}
	\begin{algorithmic}[\ref{Algo_UAV_SubSet_Possible_Location}]
		\renewcommand{\algorithmicrequire}{\textbf{Input:}}
		\renewcommand{\algorithmicensure}{\textbf{Output:}}
		\REQUIRE {Number of preferred actions $N_{\mathcal{A},s}$, $\mathcal{A}_{k}$, the antenna gain $G_r$, the target's position $\mathbf{x}_{k+H}$}
		\ENSURE  {$\mathcal{A}^{s}_{k}$}
		\FOR {$l=1:N_{\mathcal{A},k}$}
		\STATE Get $\mathbf{u}^{l}_{k+H} \in \mathcal{A}_{k}(l)$ 
		\STATE Calculate $G_r^{l} = G_r(\mathbf{x}_{k+H}, \mathbf{u}^{l}_{k+H})$
		\ENDFOR
		\STATE $\mathcal{A}_{k}^{s} = \mathcal{A}_{k}(G_r^{l} \geq \textbf{ Top } N_{\mathcal{A},s} \text{ of } G_r)$
	\end{algorithmic} 
\end{algorithm}

Following the above implementation approach, UAV motion includes two modes: \textit{i)} changing its heading angle while hovering; and \textit{ii)} moving forward to its direct location. In one planning procedure with $N_p$ cycles, the UAV needs $\lfloor |\triangle\theta|/\theta_{max} \rfloor$ cycles to rotate, and spends the remaining cycles $N_p -\lfloor |\triangle\theta|/\theta_{max} \rfloor$ to move forward without changing its heading. Here $\lfloor\cdot\rfloor$ and $|\cdot|$ are the floor and absolute operator respectively, and $\theta_{max}$ is the UAV maximum rotation angle in one cycle . The sign of $\triangle\theta$ decides the rotation direction ($+$ for the clockwise, and $-$ for the counter-clockwise).

\section{Simulation Experiments}
Implementing on a real system is time-consuming and difficult. Hence, we want to validate our systems first through several simulation experiments to: \textit{i)}  verify our tracking and planning algorithms; \textit{ii)}  
investigate how our planning parameters such as different $\alpha$ values of the R\'{e}nyi divergence or the number of discrete actions $N_{\mathcal{A},s} = |\mathcal{A}^{s}_k|$ created in Alg. \ref{Algo_UAV_SubSet_Possible_Location} contribute to the overall algorithm performance; and \textit{iii)}  compare our proposed  R\'{e}nyi divergence based planning technique with other well-known methods, and the impact of the look-ahead horizon parameters on computational time and localization accuracy. All of the simulation experiments were processed on a PC with an Intel(R) Core(TM) i7-6700 CPU @ 3.40GHz, 32GB RAM and MATLAB-2016b. 
\subsection{Tracking and Planning Simulation} \label{subsec_SIM_Target_Tracking}
This simulation was implemented to validate our approach under a synthetic scenario where all parameters (\eg,  velocity of the UAV $v_u$) are set to those expected in practice. 
In this experiment, the UAV attempted to search and localize 10 moving targets randomly located in an area of $ 500~\text{m} \times 500~\text{m}$. The following are the list of parameters used in this simulation: the sampling time step is $1$ second since the tag emits pulse signals every $1$ second. The solver performed a planning procedure every $N_p = 5$~s, and the look-ahead horizon parameters: $H =N_H t_{p} = 5$~s with the number of horizons $\ N_H = 1$ and  the planning interval $\ t_{p} = 5$~s. The UAV started from its home location at $u_{1} = [0,0,20,0]^{T}$~m, moved under the constant velocity $v_u = 5$~m/s with its maximum heading rotation angle $\theta_{max} = \pi/6$~rad/s. The number of particles for each target was capped at $N_s = 10,000$, with the number of future sample measurements $M_s = 50$, the R\'{e}nyi divergence parameter $\alpha = 0.5$, the number of actions $N_{\mathcal{A},s} = 5$. In addition, a target is considered localized if its location uncertainty, determined by the determinant of its particles covariance, is small enough---\(N_{Th} = 10,000~\text{m}^{2N_s}\) 
was chosen as the limit. The \textit{LogPath} measurement model with $P^{d0}_r = 7.7~ \text{dBm}, n = 3.1, \sigma_P = 4.22~\text{dB}$ was used to verify our proposed algorithm. To demonstrate that our algorithm was able to localize mobile targets, a \textit{wombat}---an animal that usually wanders around its area was considered. Hence, a \textit{random walk} model was used to describe its behavior with a  single target's transitional density:
\begin{align} \label{eq_dynamic_model}
	p(\mathbf{x}_k|\mathbf{x}_{k-1}) = \mathcal{N}(\mathbf{x}_k;\mathbf{Fx}_{k-1},\mathcal{Q}) 
\end{align}
where $F = \mathbf{I}_{3}$ with $\mathbf{I}_n$ is $n \times n$ identity matrix   , $\mathcal{Q} = \sigma^2_{Q}\diag([1,1,0]^T)$, $\sigma_Q = 2$~m/s. 

\begin{figure}[!tbp] 
	\centering
	\includegraphics[width=12cm]{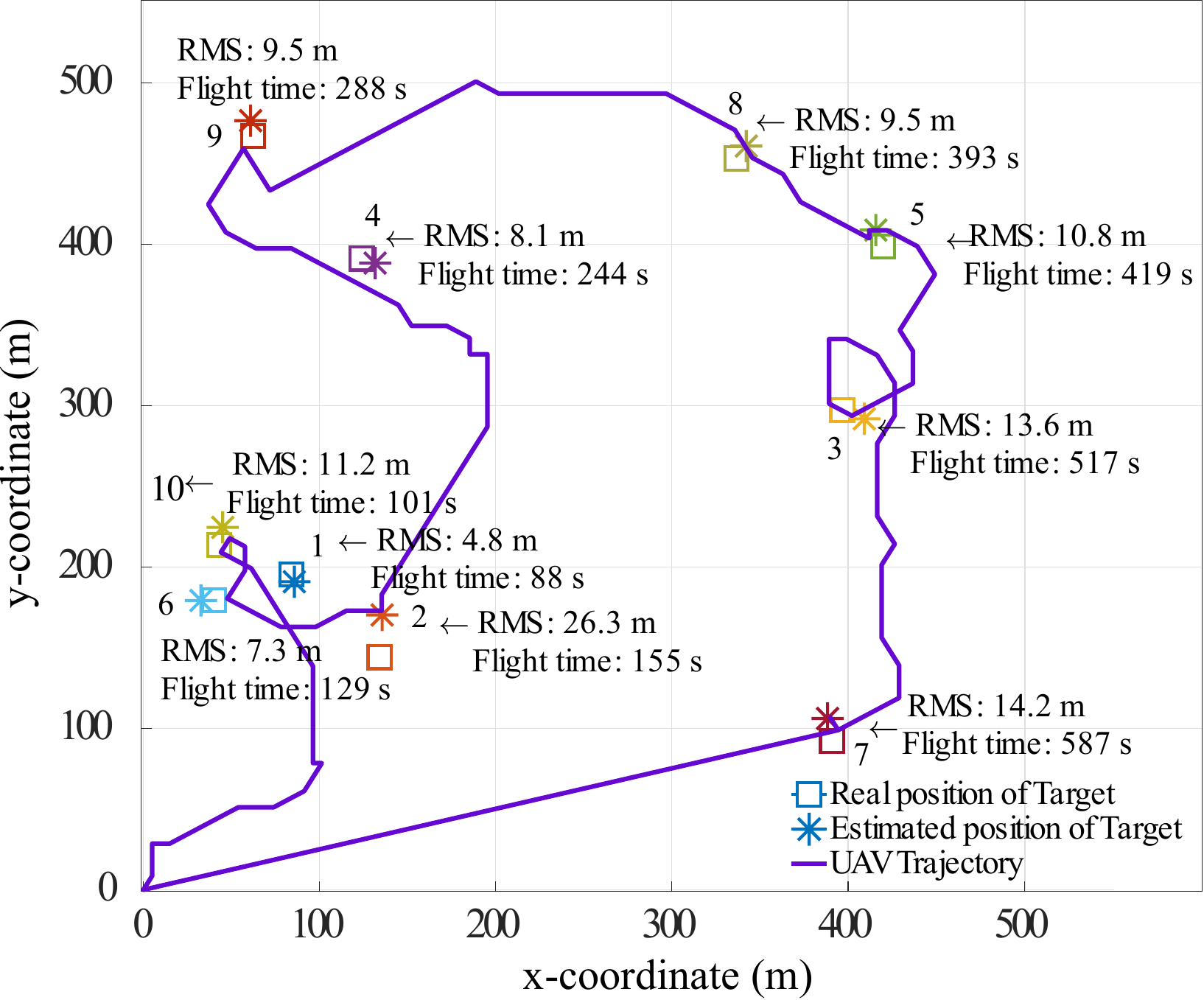}
	\caption{Simulation results with 10 mobile targets localized using a single UAV.}
	\label{fig_SIM_Estimate_Postion}
\end{figure}

Fig. \ref{fig_SIM_Estimate_Postion} shows localization results for 10 mobile targets where the estimation details are annotated next to the target's position with two indicators: \textit{Root Mean Square (RMS)}  and \textit{flight time}---see  Sec. \ref{sec_5c_MC} for definitions. In summary, for this scenario, it took the UAV \textbf{587} seconds to localize all ten moving targets at a maximum error distance of less than 15~m, except for an outlier, target \#2 (RMS = 26.3~m). At flight time 587~s, after localizing the last target (target \#7), the UAV was sent a command to fly back to its original home location. In this case, the total UAV travel distance was 1.93~km. The results demonstrate that our algorithm can search and accurately localize multiple numbers of targets in real time (about 10 minutes) and the travel distance 1.93~km is well within the capacity of commercial off the shelf drones under the 2~kg mass category.  

\subsection{Monte Carlo simulations} \label{sec_5c_MC}
For this experiment, all of the Monte Carlo setup parameters were kept the same as in Sec. \ref{subsec_SIM_Target_Tracking}, except for those under investigations. In addition, to ensure that the results were not random, all of the conducted experiments were performed over 100 Monte Carlo runs. The tracking algorithm was evaluated based on the following criterion:
\begin{itemize}
	\item \textit{Estimation Error} is the absolute distance between ground truth and estimated target location \(\mathcal{D}_{rms} = \sum_{j=1}^{N_{tg}} d^j_{rms}/N_{tg} \) with  \(d^j_{rms} = [(x^j_{truth} - x^j_{est})^2 + (y^j_{truth} - y^j_{est})^2]^{1/2} \). 
	\item \textit{Flight time} (s) for UAV to localize \textbf{all} of the targets and this includes hovering time when the UAV waits for commands from the solver to take an action. 
	\item \textit{UAV travel distance:} the total distance traveled by the UAV to track and locate all of the targets to the required location uncertainty bound; i.e the determinant of covariance being adequately small---\(N_{Th} \leq 10,000~\text{m}^{2N_s}\) .
	\item \textit{Computational cost}: We evaluate the computational cost in terms of two components: \textit{i)} execution time for the solver to execute the tracking algorithm only (called \textit{non-planning time}), and \textit{ii)} the execution time for the solver to select the best action---planning step---as well as complete the tracking task (called \textit{planning time}). 
\end{itemize}
First, our search and localization algorithms were evaluated using different $\alpha$ values for R\'{e}yni reward function in \eqref{ReyniDivergence}. Table \ref{table_MC_alpha} presents the Monte Carlo results for $\alpha = \{0.1, 0.5, 0.9999\}$. In general, the $\alpha$ value does not significantly impact the overall performance. However, applying $\alpha = 0.1$ provides the best localization results in terms of estimation error and search duration. Applying $\alpha = 0.5$ proposed in \cite{Ristic2010,ristic2010information} results in the worst performance, it increases flight time and travel distance necessary to complete the localization task. Using $\alpha = 0.9999$ (considered as using KL divergence which is a popular information gain measure) helps to save UAV travel distance while sacrificing location accuracy. One explanation for this scenario is that our noisy measurement causes the predicted posterior $p(\mathbf{x}_{k+H}|\mathbf{z}_{1:k},\mathbf{z}^{(m)}_{k+1:k+H}(\mathbf{a}))$ in \eqref{ReyniDivergence} to be less informative due to high uncertainty. Therefore, the reward function should place more emphasis on the current posterior instead by using a small $\alpha$ value or setting $\alpha \rightarrow 1$ to completely ignore the future posterior. This also explains the reason for the worst localization performance observed when $\alpha = 0.5$ (equally weighting the current and the future posterior).

\begin{table}[!b] 
	\centering
	\caption{Localization performance for different alpha values.}
    \vspace{0.2cm}
	\begin{tabular}{l |r r r} 
		\cmidrule{2-4} 
		& $\alpha = 0.1$ & $\alpha = 0.5$ & $\alpha = 0.9999$ \\ 
		\midrule\midrule
		RMS (m)               
		& \textbf{12.35} & 12.77 & 12.96 \\ 
		Flight time (s)  & \textbf{724}  & 741 & 727 \\
		UAV travel distance (km) & 2.38 & 2.41 & \textbf{2.34}\\
		\hline
	\end{tabular}
	\label{table_MC_alpha}
\end{table}

Second, we conducted experiments to understand how the number of actions $N_{\mathcal{A},s}$ created by Alg. \ref{Algo_UAV_SubSet_Possible_Location} affects our tracking performance in term of planning time and localization error. 
Table \ref{table_MC_action} shows Monte Carlo results for $N_{\mathcal{A},s} = \{2,3,4,5,6,7\}$.  Increasing the number of actions beyond four does not necessarily lead to better planning decisions because of the directionality of the antenna gain. Since the antenna gain is not omnidirectional, some actions result in changing the heading where antenna gain along the propagation path between the UAV and the target is lower; when the number of actions evaluated is increased, we encounter instances when an action leading to such a lower antenna gain, in fact, results in a higher reward. This result is a consequence of the inherent uncertainties in the models used in tracking and planning. Thus,  we can see that $N_{\mathcal{A},s}=4$  provides an adequate pool of actions to yield the best localization performance in terms of estimation error, flight duration, and travel distance; a desirable result for realizing real-time planning with limited computational resources.
\begin{table}[!tbp] 
	\centering
	\caption{Localization performance for different number of actions.}
    \vspace{0.2cm}
	\begin{tabular}{l |  *{6}{r}} 
		\hline
		Number of actions $N_{\mathcal{A},s}$ & 2 & 3 & 4 & 5 & 6 & 7 \\ 
		\midrule\midrule
		RMS (m)               
		& 14.18 & 12.64 &\textbf{12.17}  & 12.27 & 12.83 & 12.63\\ 
		Flight time (s)  & 840 & 781 & \textbf{693} & 723 & 756 & 799 \\
		UAV travel distance (km) & 2.62 & 2.53 & \textbf{2.39} & 2.50 & 2.52 & 2.70\\
		Planning time (s) & \textbf{1.16} & 1.19 & 1.23 & 1.27 & 1.36 & 1.47\\
		\hline
	\end{tabular}
	\label{table_MC_action}
\end{table}

\begin{figure}[!tbp] 
	\centering
	\includegraphics[width=10.0cm]{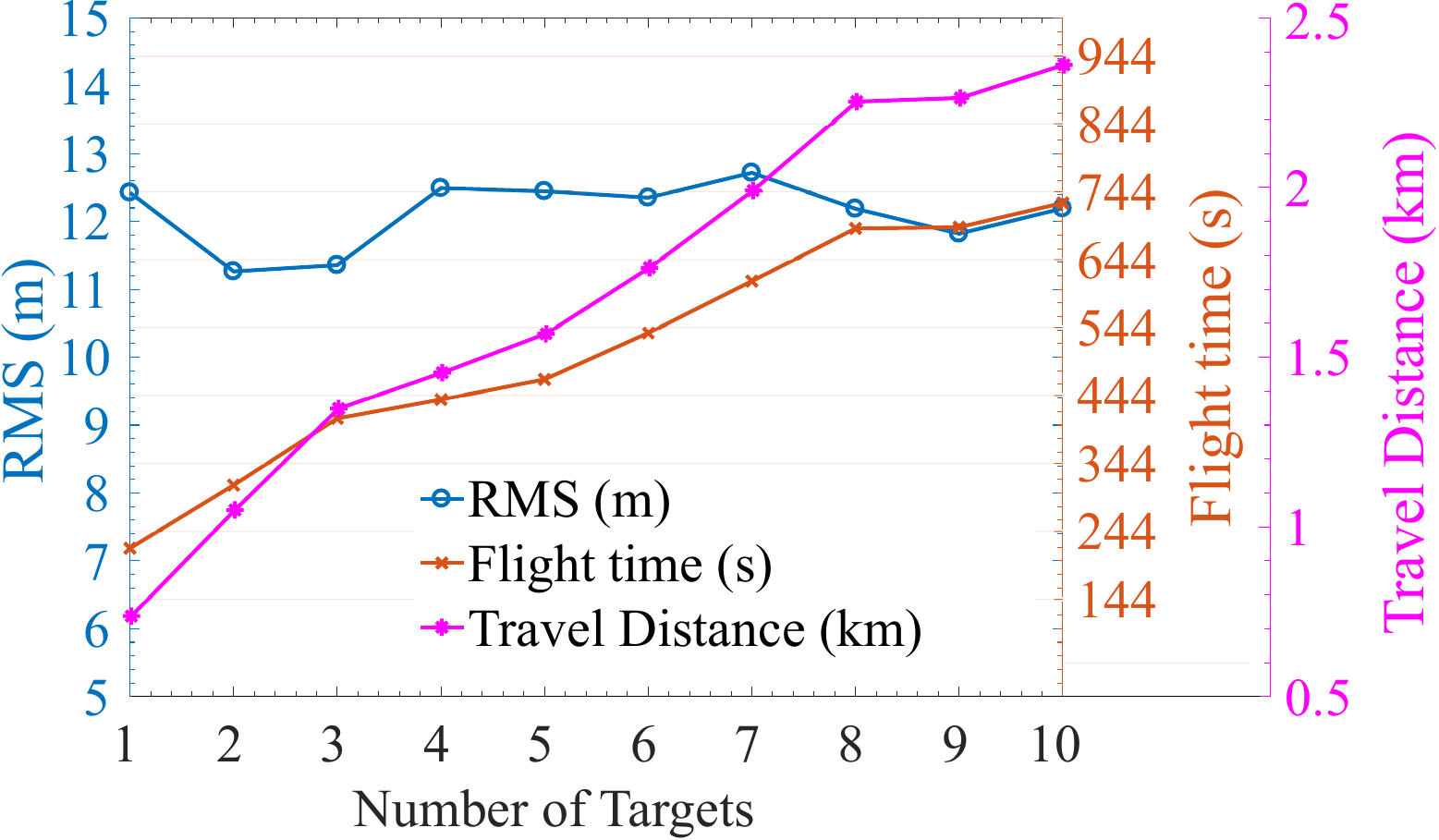}
	\caption{Localization performance for different number of targets $N_{tg}$ increase from 1 to 10.}
	\label{fig_MC_Target_Change_Comparison}
\end{figure}

Third, we want to examine the performance of  our proposed algorithm under an increasing number of targets; in this study, we increase the maximum number of targets \(N_{tg}\) from 1 to 10. As depicted in Fig. \ref{fig_MC_Target_Change_Comparison}, our algorithm's estimation error was stable and invariant to the number of targets. Moreover, it is reasonable that the flight time and the travel distance increased linearly with target numbers because it took more time and power to track more targets. 

Fourth, we examined the performance of the  information gain measure, R\'{e}nyi divergence, under different look-ahead horizons $H = N_{H} t_p$ compared to: \textit{i)} Shannon entropy \cite{cliff2015online}; \textit{ii)} a naive approach that moves UAV to the closest estimated target location; and \textit{iii)} a uniform search with the predefined path used in \cite{ristic2010information}. Table \ref{table_MC_planning} shows the Monte Carlo comparison results among various planning algorithms. All the parameters were reused from the Sec. \ref{subsec_SIM_Target_Tracking}, except for $\alpha = 0.1$ and $N_{\mathcal{A},s} = 4$ were updated based on the previous experimental results. 
The results demonstrated that R\'{e}nyi divergence based reward function leads to significantly better planning strategies in comparison with other reward functions in terms of localization accuracy, including Shannon entropy with the same horizon settings ($N_H = 1;t_p = 5$). 
For R\'{e}nyi reward function itself, the large look ahead horizon number $N_H > 1$  helps to improve the localization accuracy; however, it requires higher computational power (planning) and causes the UAV to travel further. Using $N_H = 1;t_p = 5$~s provides the best trade-off between computational time and accuracy.

\begin{table}[!tbp] 
	\centering
	\caption{Localization performance for different planning algorithms.}
    \vspace{0.2cm}
	\begin{tabular}{L{3cm} | *{2}{R{2cm}} R{2.4cm} | *{4}{R{1cm}} } 
		\cmidrule{2-8}
		& \textbf{Uniform} & \textbf{Closest Target}  & \textbf{Shannon} \cite{cliff2015online} &  \multicolumn{4}{c}{\textbf{R\'{e}nyi}} \\[0.5ex]
		\midrule\midrule 
		$N_{H}$            & N/A & N/A & 1 & 1  & 3 & 5 & 10  \\
		$t_p$ (s)          & N/A & N/A & 5  & 5 & 1 & 1 & 1  \\
		\hline
		RMS (m)                   & 18.8 & 13.4 & 12.6 & 12.5 & 12.4 & 12.0 & \textbf{11.6}  \\
		Flight time (s)   & 921 & 799 & 774 & \textbf{699}& 889 & 811 & 822  \\
		UAV travel distance (km)           & 3.72 & 2.29 & 2.54 & \textbf{2.27} & 2.99 & 2.82 & 2.42  \\
		Planning Time (s)             & 1.58 & \textbf{1.11} & 1.38 & 1.28  & 1.53 & 1.65 & 2.71  \\
		Non-planning Time (s)         & 1.58 & 1.03 & 0.99 & 0.97& \textbf{0.96} & 0.97 & \textbf{0.96}  \\
		\hline
	\end{tabular}
	\label{table_MC_planning}
\end{table}

\textbf{Summary:} According to the above simulation results, we select $\alpha = 0.1$, $N_{\mathcal{A},s} = 4$, and $N_H = 1, t_p = 5 ~s$ as the planning parameters for the field experiment since these parameters provides the lowest computational cost, best performance in term of location estimation error, travel distance and flight time.
\section{Field Experiments}

We describe here our extensive experiments regime to validate our approach and evaluate the performance of our aerial robot system in the field. Our aim is to: \textit{i)}  investigate the possibility of signal interference from spinning motors of a UAV on RSSI measurements; \textit{ii)} estimate the model parameters in the sensor model and validate the proposed model; and \textit{iii)} conduct field trials to demonstrate and evaluate our system capabilities. 

\subsection{Rotor noise} 
\begin{figure*}[!tbp] 
	\centering
	
	\includegraphics[clip, width=14cm]{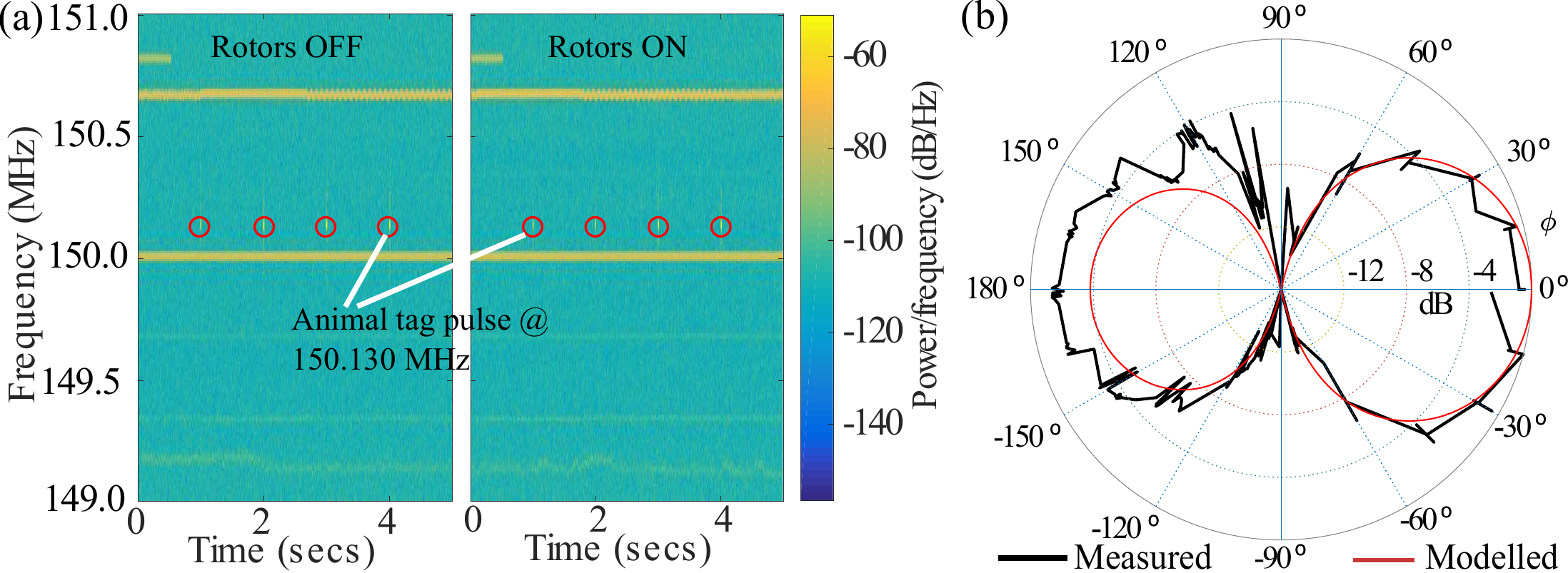}
	
	\caption{a) Waterfall plot for the rotor noise experiment when four motors spun at maximum rotation speed. b) Normalized antenna gain in E-plane $G(\phi)$. The red line is gain modeled pattern and black line is the normalized measured gain pattern from 30 measurements collected by rotating the UAV heading at 15$^\circ$ intervals.}
	
	\label{Combine_RotorNoise_AntennaGain_MeasureModel}
\end{figure*}

We investigated the rotor noise to confirm that our system is not affected by the electromagnetic interference from the UAV's motors. It also helps to clear the concern raised in~\cite{cliff2015online} that the rotor noise may affect the RSSI measurements. Four motors of the 3DR IRIS+ quad-copter shown in Fig.\ref{fig_UAV_System_Overview} were used in this experiment. The RSSI data of a radio collar were measured across 149 MHz to 151 MHz frequency spectrum when four motors were operating at $20\%$, $ 50\%$, $100\%$ of its maximum speed of $10,212$ revolutions per minute. Fig \ref{Combine_RotorNoise_AntennaGain_MeasureModel} (a) shows the frequency spectrum of the received signal. We can see that there was no difference in the frequency characteristics when the rotors were in ON and OFF states. This result confirms that the rotors do not spin fast enough to generate high-frequency interference to impact our RSSI measurements.

\subsection{Sensor model validation and parameter estimation}\label{sec_emperical_meas}
\textbf{Antenna Gain:}
The antenna gain pattern was measured to verify its directivity compared to the antenna gain model $G_r(\mathbf{x,u}) = G_r(\phi)$ calculated---following~\cite[p.1252]{orfanidis2002electromagnetic}---based on the physical design as discussed in Sec. \ref{sec_Antenna_Gain}. Fig. \ref{Combine_RotorNoise_AntennaGain_MeasureModel}b shows the measured and modeled radiation patterns  $G_r(\phi)$ in the E-plane. In the measurement process,  $\phi$ is evaluated as the angle between the UAV heading, changed through $0^{\circ}$ to $360^{\circ}$, and the direction from its position to a fixed location of a VHF radio tag. 
The result shows that the front-to-back ratio is smaller (2~dB) than expected and this is an artefact of folding the reflector on our design.  

\begin{figure}[!tbp] 
	\centering
	\includegraphics[width=10cm]{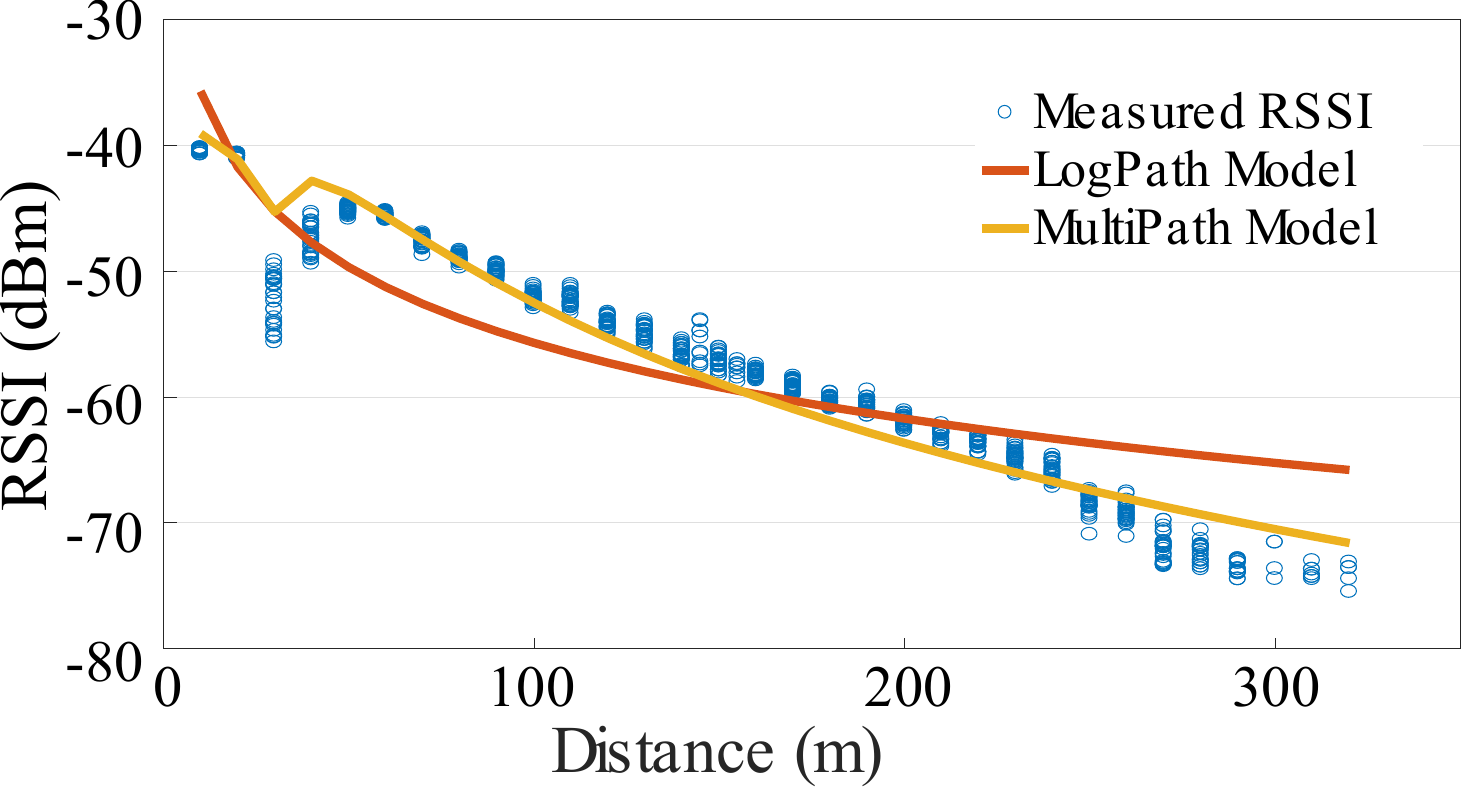}
	
	\caption{Plot of measured RSSI data points and its model estimates over a distance from 10~m to 320~m at 10~m intervals.} 
	
	\label{fig_Meas_Model}
\end{figure}

\textbf{Signal propagation model parameter validation:}
We collected RSSI data points over a range from 10~m to 320~m between the UAV and a VHF radio tag. 
The tag and the UAV were kept at a height of 5~m  above ground during this experiment. The tag was stationary at all times, while the UAV was directed to move away in a straight line from the tag at 10~m intervals whilst hovering at each location to allow the collection of approximately 30 measurements. The UAV heading was maintained to ensure consistent antenna gain during the experiment. Since we operated in an open terrain over a grassland, we selected the path loss exponent $n=2$ suitable for modelling free space path loss. Fig. \ref{fig_Meas_Model} shows the measured RSSI and the propagation models obtained using a nonlinear regression algorithm to estimate model parameters; we have the following results for reference power $P_r^{d0}$ in \eqref{eq_LogPath}, \eqref{eq_MultiPath} at the reference distance $d_0 = 1$ m,  and  measurement noise variance $\sigma_{P}$ in \eqref{eq_pathloss}: 

\begin{itemize}
	\item \textbf{LogPath model :} $P_r^{d0} = -15.69 ~\text{dBm}$ ; $\sigma_{P} = 4.21$~dB.
	\item \textbf{MultiPath model :} $P_r^{d0} = -15.28 ~\text{dBm}$ ; $\sigma_{P} = 2.31$~dB.
\end{itemize}

The results show that both models, as expected, derived a similar reference power $P_r^{d0}$ whilst providing a reasonable fit to measurement data. This affirms the validity of our propagation model. Although \textit{LogPath} model is reasonable, \textit{MultiPath} model is more accurate and yields a smaller measurement noise variance. The results confirm the impact of ground reflections, especially close to the signal source.  

\subsection{Field Trials} \label{sec_field_trials}
We designed and conducted two sets of field experiments that included 20 autonomous missions as described below.
\begin{itemize}
\item \textbf{First set of trials (autumn season):} We conducted a total of 16 autonomous flights with two mobile radio-tags to evaluate the measurement models and demonstrate the robustness of our system (see Section~\ref{sec:first-trial}). 
\item \textbf{Second set of trials (winter and wet season):}  We conducted 4 autonomous flights with the best performing measurement model.  These experiments were aimed at demonstrating the multi-target tracking capability of our aerial robot platform under a mix of stationary (3 radio-tags) and mobile (2 radio-tags) target dynamics. In particular, we subjected our system to two highly mobile targets. Notably, these trials were conducted during the wet winter months when the test zone was over-grown with grass and shrubs. Therefore, these experiments demonstrate our system's capability to plan a trajectory to track multiple radio-tagged objects with differing motion dynamics and under different environmental settings (see Section~\ref{sec:second-trial}).
\end{itemize}

Our experiments were designed around the University of Adelaide and CASA (Civil Aviation Safety Authority, Australia) regulations governing the conduct of UAV research. Given the need to operate in an autonomous mode, our flight zone, as well as the scope of the experiment, was restricted to University-owned property designated for UAV flight tests. Prior to gaining ethical and regulatory clearances to progress our field trial to a wildlife species of interest to conservation biologists, our first objective is to evaluate and demonstrate a robust working prototype. This is a necessary condition to gain both regulatory and ethical approval. Further, it is not feasible to have a wildlife species of interest at the remote test site and conduct experiments to systematically evaluate the aerial robot system. Therefore we chose to conduct experiments with human test subjects with stipulated safety measures in an area allocated for field tests. This allowed us to create various target motion dynamics as well as obtain accurate ground truth data for tag locations to evaluate our system. Notably, our measurement model is based on the Received Signal Strength Indicator (RSSI-based) measurements of signals transmitted from radio-tags. Hence, there is no technical difference whether the radio-tags are carried by humans or wildlife.  

\begin{figure}[!tbp] 
	\centering
	\includegraphics[width=6cm]{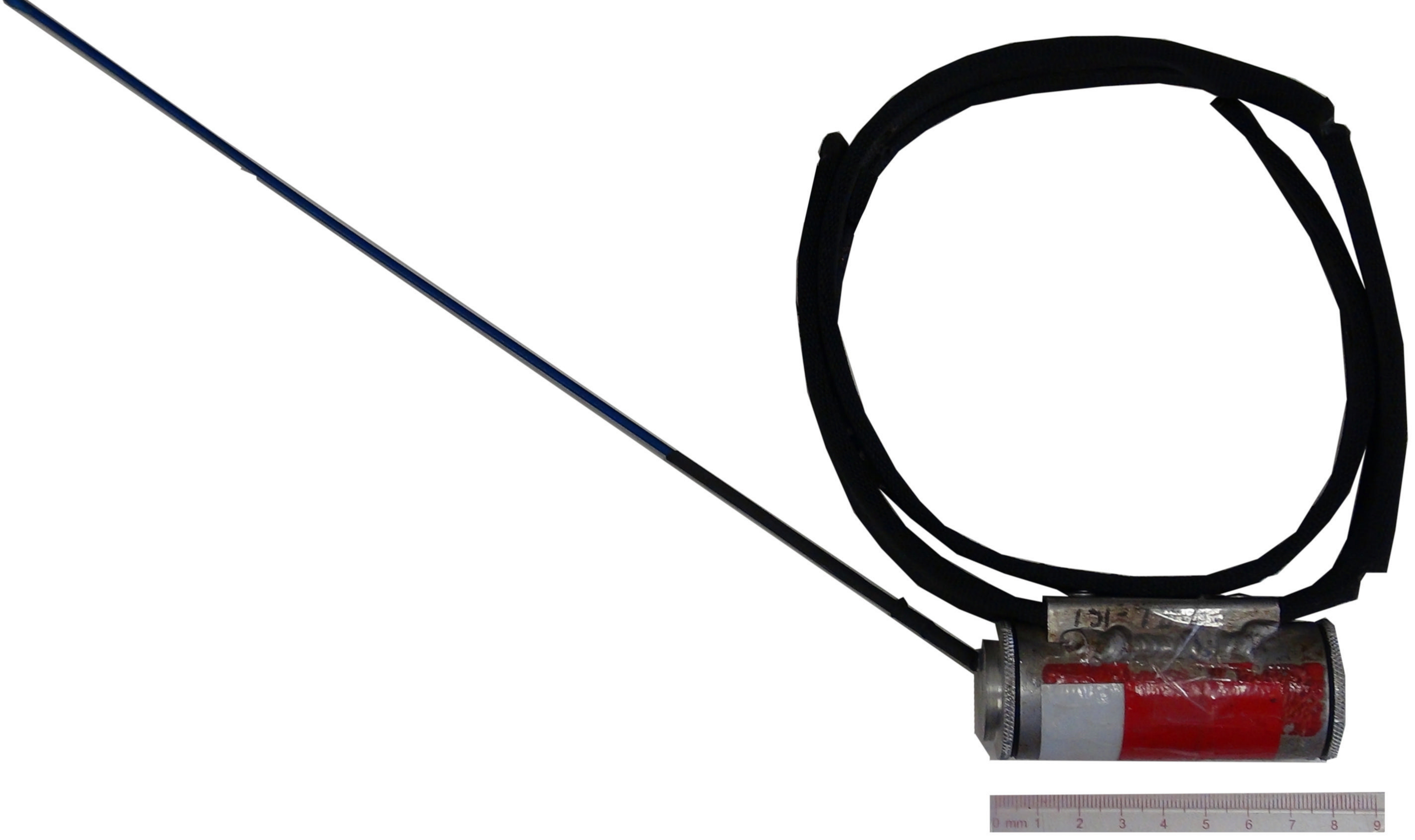}
	\caption{A collar used for radio tagging the endangered species of Southern Hairy Nose (SHN) wombats used in our field experiments. Each tag used in our experiments transmits an unmodulated on-off-keyed signal with a pulse width ranging from $10-20$~ms, at a period of approximately $1$~s, and using a unique frequency in the range of $150-152$~MHz.}
	\label{fig_radio_tag}
\end{figure}

In the field trials, the task of the aerial robot system was set to search and localize radio-tags undergoing various motion dynamics in a search area $75~\text{m} \times 300~\text{m}$ ($2.25$ hectares).
Instead of wildlife, we relied on volunteers to wear a VHF radio tag of the type shown in Figure~\ref{fig_radio_tag} on their forearm, and carry a mobile phone-based GPS data logger in their hands to obtain ground truth data. We were required to have two extra personnel stationed to maintain constant sight of the UAV as well a pilot with RePL (Remote Pilot License) in the field capable of aborting the autonomous mode and transferring control to manual operations mode. 

\subsection{First Set of Trials}\label{sec:first-trial}

In this section, we present the first set of field trials to demonstrate the planning method for tracking mobile targets. We also compare localization performance between the two signal propagation models:  \textit{LogPath} model and \textit{MultiPath} model derived in section \ref{sec_emperical_meas}. We used two VHF radio collars for these trials.

\begin{figure}[!tbp] 
	\centering
		\includegraphics[clip,width=1\textwidth]{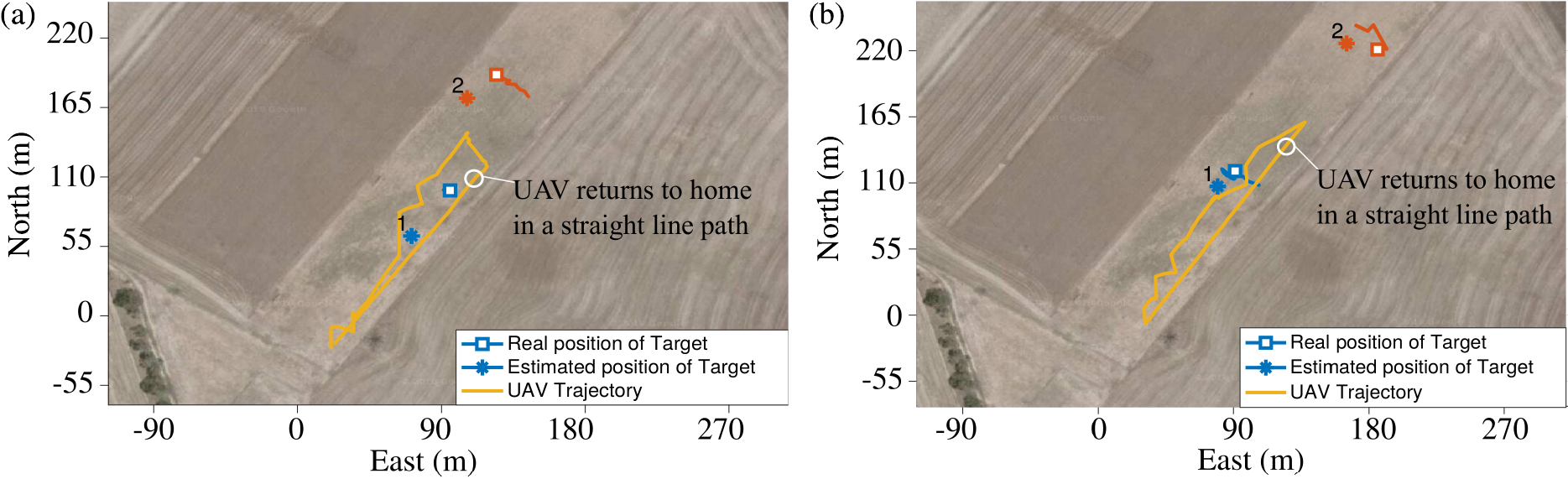}
	\caption{Field experiment results to search, track and localize two mobile tags for the two different measurement models. a) Standard \textbf{LogPath}. b) \textbf{MultiPath}.}
	\label{fig_Field_Experiment_Result}
\end{figure}

Fig. \ref{fig_Field_Experiment_Result} shows the tracking and localization results along with UAV trajectories based on the two different measurement models. As expected, we observe the UAV planning has a tendency to approach the target's position since when the distance between the UAV and targets reduces, the RSSI measurement uncertainty is reduced. thus it helps to reduce the uncertainty and increase the information gain. We can observe a clear difference in the \textit{LogPath} model and \textit{MultiPath} model where UAV pursues the second target after completing the tracking task for target 1. The more accurate \textit{MultiPath} model is able to track and localize the second target without needing a close approach. We can also observe that using \textit{LogPath} model, where multipath propagation is not modeled but is clearly dominant close to the target, leads to a poorer localization accuracy despite the path planning algorithm leading the UAV close to the target. 

\begin{figure}[!tbp] 
	\centering
	\includegraphics[width=1.0\textwidth]{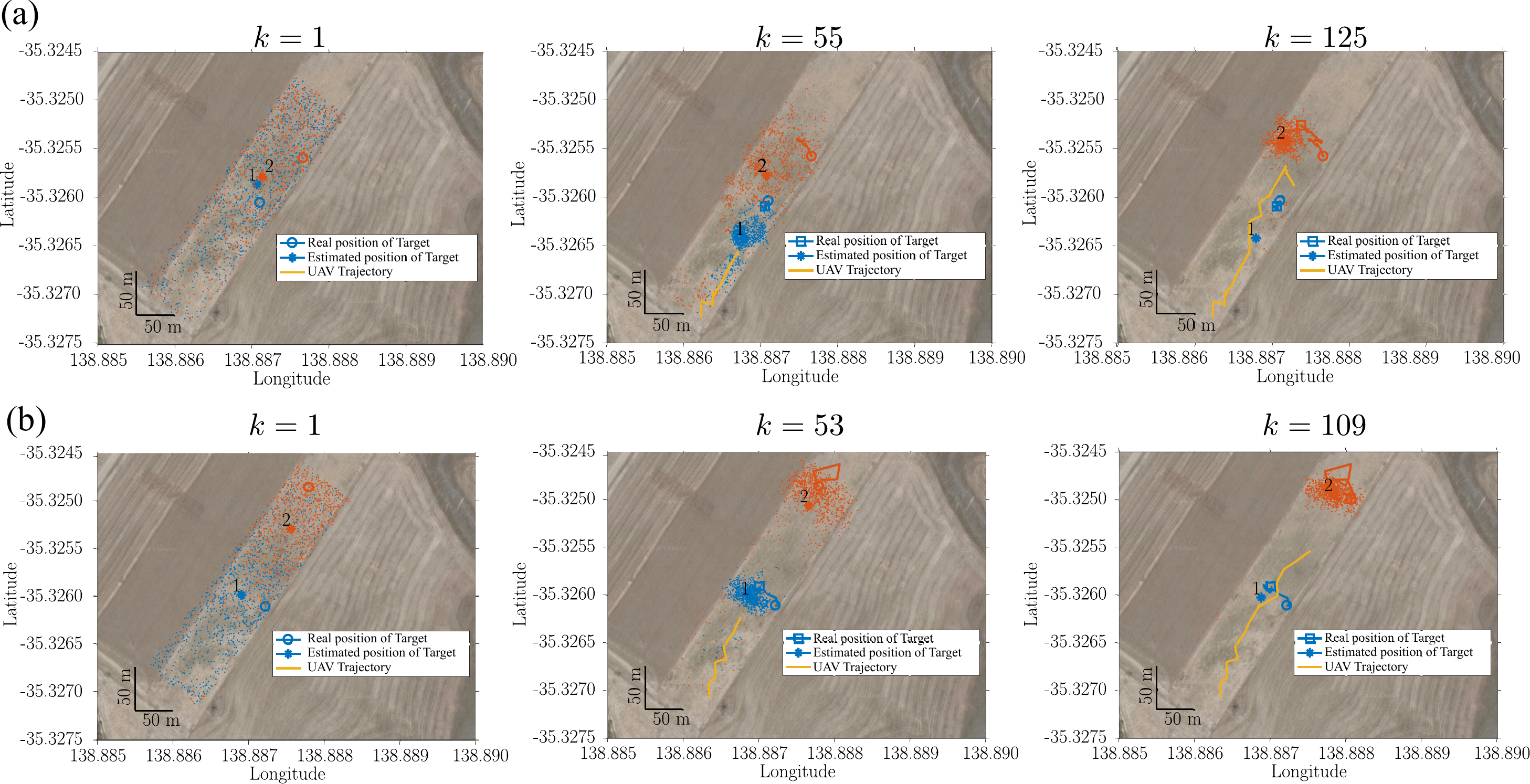}
	\caption{The intermediate distributions of belief density representing the location of the radio-tags for the two scenarios in Fig.~\ref{fig_Field_Experiment_Result}. Here, Fig~\ref{fig_Field_Experiment_Particle}a demonstrates the convergence of the belief density of the radio-tag positions using the standard \textbf{LogPath} measurement model in Fig.~\ref{fig_Field_Experiment_Result}a) after first observation ($k=1$), tag 1 is localized ($k=55$), and tag 2 is localized ($k=125$). Similarly,  Fig~\ref{fig_Field_Experiment_Particle}b demonstrates the convergence of the belief density of the radio-tag positions using the \textbf{MultiPath} measurement model in Fig.~\ref{fig_Field_Experiment_Result}b) after the first observation ($k=1$), tag 1 is localized ($k=53$), and tag 2 is localized ($k=109$). The blue and orange dots represent the starting positions of tag 1 and tag 2, respectively. The square symbols denote the ground truths of the localized tags; the star symbols denote the estimated positions of the tags. The solid yellow lines represent the UAV trajectories.}
	\label{fig_Field_Experiment_Particle}
\end{figure}

Fig. \ref{fig_Field_Experiment_Particle} shows the particle distribution after the first observation is updated and when the targets are tracked and localized using the two measurement models. We can see that the solver is able to estimate the two tag positions quiet accurately even after the update using the initial observation; however, the uncertainty (as noted by the particle distribution) is still very high. Interestingly, \textit{MultiPath} model location uncertainty is significantly less where target 1 is placed in the bottom half of the field while target 2 is placed in the top half of the field. Target 1, being closer to the UAV, is localized first, with under $55$ measurements for both measurement models. At the time when target~1 is localized, the uncertainty of target~2 is relatively higher for the \textit{LogPath} model. The \textit{MultiPath} model required significantly fewer measurements to track and localize target~ 2. As expected, both measurement models required significantly more measurements to localize the second target given the high measurement uncertainty associated with being much further than the first target from the UAV during its flight. Furthermore, random walk of the second target provided a challenging scenario since target~2 typically moved a larger distance around the field compared to the random walk performed by target~1. 

Although the solver guides the UAV to move toward a target's position in both measurement models, as expected, the standard \textit{LogPath} model is less accurate compared to the \textit{MultiPath} model shown in Fig. \ref{fig_Meas_Model}; thus, the uncertainty when using the \textit{LogPath} model is higher and leads to longer time duration to localize the two tags. Albeit model uncertainty, the \textit{LogPath} model is still capable of locating both moving targets within the flight time capability of the UAV. The consequence of model uncertainty resulting from the  \textit{LogPath} model is more apparent when the UAV makes an approach to the target and the distance to the target is less than $50$~m. This is evident in comparing the belief density in Fig.~\ref{fig_Field_Experiment_Particle}a at $k=125$ to that in Fig.~\ref{fig_Field_Experiment_Particle}b at $k=109$. We can see that the target location uncertainty increase for the \textit{LogPath} model in the vicinity of 50~m and as a result, the UAV requires an increased number of maneuvers to track and locate the target.

Table \ref{table_Field_Experiment_Result} presents the summary comparison results of location estimates between the two measurement models. Smaller RMS (root mean square) estimation error values suggest a higher accuracy,  while shorter flight times and travel distance to localize all targets are highly desirable for a practicable system given the power constrained nature of commodity UAVs. The results confirm that the \textit{MultiPath} model is superior to the standard \textit{LogPath} model since it has been able to account for ground reflections. Further, the UAV is not required to approach the target closely to reduce its measurement uncertainty when using the \textit{MultiPath} model. 

The results in Table \ref{table_Field_Experiment_Result} also demonstrate that our proposed method can localize two \textit{mobile} targets with a shorter flight time and better accuracy compared to the method in \cite{cliff2015online}. The RMS flight time realized with the \textit{MultiPath} model is one-sixth of that in \cite{cliff2015online}). Although our experiments were not performed with a live target animal species of interest to conservation biologists, we search and locate two mobile radio-tagged targets. In contrast, \cite{cliff2015online} method was formulated and implemented to locate a single stationary target. However, the approach in \cite{cliff2015online} was evaluated with a stationary radio-collared live bird while our field experiments were conducted with human test subjects.

\begin{savenotes}
	\begin{table}[!tbp] 
		\centering
		\caption{Comparison of localization performance.} 
        \vspace{0.2cm}
		\begin{tabular}{L{3cm} | C{2cm} C{1cm} C{2cm} *{2}{C{3cm}}} 
			\hline
			\textbf{Model} & Target Type & Trials & RMS (m) & Total Flight Time (s) & Travel Distance (m) \\ 
			\midrule\midrule
			LogPath & Mobile & 8 & $30.1\pm 12.8$ & $255\pm104$ & $549\pm167$\\ 
			MultiPath & Mobile & 8 & $\mathbf{22.7}\pm13.9$ & $\mathbf{138}\pm53$ & $\mathbf{286}\pm121$ \\
			\cite{cliff2015online} & Stationary & 6 & $23.8\pm14.0$ & 838\footnote{ Information regarding the total flight is not reported in~\cite{cliff2015online}, however, as shown in Fig. 9 in~\cite{cliff2015online}, one observation took 76.21s and one trial needed 11 observations, hence total flight time is $11 \times 76.21 = 838.31 s$} & N/A\\
			\hline
		\end{tabular}
		
		\label{table_Field_Experiment_Result}
	\end{table}
\end{savenotes}

\subsection{Second Set of Trials}\label{sec:second-trial}

In this section, we present the second set of field trials. We use the \textit{Multipath} measurement model because it provides a better measurement likelihood as shown in the tracking accuracy and flight time results in Table~\ref{table_Field_Experiment_Result}.  
We can see from Fig.~\ref{fig_dsp_chains}, the SDR-based signal processing architecture used in our system scales to enable tracking a large number of radio tags. The number of VHF radio-tags that can be tracked and localized is only limited by the hardware, such as the battery life of the UAV and the receiver noise of the SDR. In order to demonstrate scalability and robustness, we used five radio-tags. In order to demonstrate the capability of our system to accommodate different animal behaviors, we used two highly mobile targets  (target $1,2$) and three stationary targets (target $3,4,5$). Further, to demonstrate the robustness of our measurement model,  we conducted these trials in the wet, winter season in South Australia where the test site was representative of a grassland with shrubs and moisture. 
We conducted four field missions in which the task of our aerial robot system was to track and localizes five targets as opposed to two mobile targets investigated in Section~\ref{sec:first-trial}. All other experimental settings were as described in Section .\ref{sec:first-trial}.

\begin{figure}[!tb] 
	\centering
	\includegraphics[clip,width=1\textwidth]{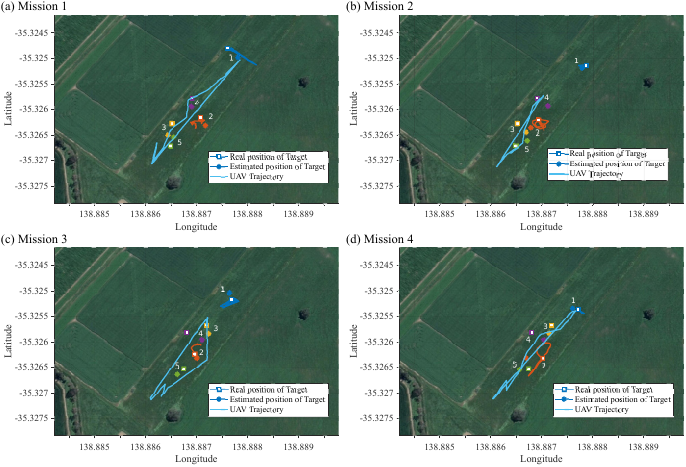}
	\caption{Four autonomous field experiment missions to search, track and localize five targets: two mobile targets (target $1,2$) and three stationary tags (target $3,4,5$). Fig.~\ref{fig_Field_Trial_5tags}(a), (b), (c), and (d) corresponds to the sequence of the missions in Table~\ref{table_localization_5tags}. The  square  symbols  denote  the  ground  truth  of  the  localized  radio-tags;  the  star  symbols  denote  the  estimated  positions  of  the  radio-tags;  the  solid  blue  lines  represent  the trajectories planned by the autonomous aerial robot  to track the set of five VHF radio-collared tags.}
	\label{fig_Field_Trial_5tags}
\end{figure}

Fig.~\ref{fig_Field_Trial_5tags} depicts the UAV and mobile target trajectories together with tracking and localization results. Table.~\ref{table_localization_5tags} presents a quantitative summary of the results from the four field missions. The results show that when the targets are highly mobile, such as target $1$ in Fig.~\ref{fig_Field_Trial_5tags}a or target $1$ and $2$ in Fig.~\ref{fig_Field_Trial_5tags}d, the UAV takes longer flight paths to be able to localize these highly mobile targets. This is because the UAV undertakes control actions to position itself to reduce measurement uncertainty. Consequently, we also see that the UAV path planning algorithm undertakes control actions to navigate the UAV closer or follow targets to quickly reduce measurement uncertainty. In contrast, when the targets are less mobile as shown in Fig.~\ref{fig_Field_Trial_5tags}b-c, the UAV can easily localize the targets with fewer measurements, shorter flight paths, and without needing to approach the targets. Thus, when targets are less mobile, the UAV requires less flight time to accurately track and localize them. We can see that our planning for tracking approach was robust with respect to various target motion dynamics we have created. Further, the results summarized in Table~\ref{table_localization_5tags} demonstrate that our localization results were consistently high across all four missions. 

As expected, our aerial robot system can successfully track and localize multiple radio tags. In relation to the first set of field trials, we can also see that our system is: \textit{i)}~scalable to a larger number of VHF radio-tags; \textit{ii)}~robust against variations in environmental conditions; and \textit{iii)}~robust with respect to various target behaviors.

\begin{table}[!tb]
\centering
\caption{Localization performance over four field missions to track and localize five radio-tagged targets.}
\vspace{0.2cm}
\label{table_localization_5tags}
\begin{tabular}{l|r|r|r|r|r|r|r|}
\cline{2-8}
                                            & \multicolumn{6}{c|}{RMS (m)}                                                                                                                                                                                                                      & \multicolumn{1}{c|}{\multirow{3}{*}{\textbf{\begin{tabular}[c]{@{}c@{}}Flight time\\ (s)\end{tabular}}}} \\ \cline{1-7}
\multicolumn{1}{|l|}{\textbf{Target dynamics}} & \multicolumn{2}{c|}{\textit{Mobile}}                                      & \multicolumn{3}{c|}{\textit{Stationary}}                                                                        & \multicolumn{1}{c|}{\multirow{2}{*}{\textbf{Mean}}} & \multicolumn{1}{c|}{}                                                                                    \\ \cline{1-6}
\multicolumn{1}{|l|}{\textbf{Target \#}}       & \multicolumn{1}{c|}{\textbf{Target 1}} & \multicolumn{1}{c|}{\textbf{Target 2}} & \multicolumn{1}{c|}{\textbf{Target 3}} & \multicolumn{1}{c|}{\textbf{Target 4}} & \multicolumn{1}{c|}{\textbf{Target 5}} & \multicolumn{1}{c|}{}                               & \multicolumn{1}{c|}{}                                                                                    \\ \hline
\multicolumn{1}{|l|}{\textbf{Mission 1}}    & 27.3                                & 19.1                                & 27.2                                & 18.1                                & 19.9                                & 22.3                                                & 163                                                                                                      \\ \hline
\multicolumn{1}{|l|}{\textbf{Mission 2}}    & 9.3                                 & 21.8                                & 24.9                                & 25.4                                & 23.7                                & 21.0                                                & 143                                                                                                      \\ \hline
\multicolumn{1}{|l|}{\textbf{Mission 3}}    & 15.0                                & 9.3                                 & 18.7                                & 30.6                                & 16.3                                & 18.0                                                & 128                                                                                                      \\ \hline
\multicolumn{1}{|l|}{\textbf{Mission 4}}    & 10.0                                & 29.6                                & 18.4                                & 25.1                                & 16.6                                & 19.9                                                & 165                                                                                                      \\ \hline
\end{tabular}
\end{table}

\section{Discussion}
In this section, we summarize and discuss results from our approach as well as compare and discuss our results in the context of the recent study by  \cite{cliff2015online} (see Section~\ref{sec:compare-cliff}). We then reflect upon the lessons learned from our field trials to build, test and evaluate a new approach following a different school of thought for autonomous tracking and localization of VHF radio-tags (see Section~\ref{sec:lessons}). Our work, being a first, is not without limitations. We discuss these in Section~\ref{sec:limiataions}.

\subsection{Comparison}\label{sec:compare-cliff}
Table \ref{table_Field_Experiment_System} presents a complete comparison between our proposed system and \cite{cliff2015online} system. Notably, our search area is smaller compared to \cite{cliff2015online} ($75~\text{m} \times 300~\text{m}$ v.s $1000~\text{m} \times 1000~\text{m}$) due to our test flight zone restrictions; however, we have set up our initial distance from the UAV home position to its farthest target's position to be equivalent to the distance of the stationary target in~\cite{cliff2015online}; approximately 300~m. 
Although we have tried to replicate the distance to the location of a radio-tag, the detection range is determined by a number of factors other than the specification of the receiver and the antenna used. The detection range is heavily  influenced by the transmitted power of a radio-tag, which is adjusted based on application requirements and varies in different environments, even for the radio collars form the same manufacturer. 
Therefore, we have not directly compared the detection range. Instead, we have tried to achieve a similar UAV-to-target distance in our experimental settings.

In general, as shown in Table \ref{table_Field_Experiment_System}, our system is more compact, lighter,  and has a payload that is one-third of that in~\cite{cliff2015online} and consequently capable of longer flight times on any given UAV. Our total system mass being under $2$~ kg enables ecologists in jurisdictions such as Australia~\cite{casaac101}, Germany~\cite{GermanyBMVI2017} and India~\cite{IndiaDGCA2018}  
to operate our system without a remote pilot license (RePL) and regulatory burdens.
Moreover, as shown in Table \ref{table_Field_Experiment_Result}, compared to the bearing-only method requiring full rotations of a UAV at each observation point, the ability to instantly collect RSSI measurements also helps reduce flight times significantly. Furthermore, as discussed in \cite{arulampalam2002tutorial}, the computational cost for grid-based methods used in \cite{cliff2015online} increases dramatically with the number of cells whilst the grid must be dense enough to achieve accurate estimations; \eg, a grid-based filter with $N$ cells conducts $\mathrm{O}(N^2)$ operations per iteration, while a similar particle filter with $N$ particles only requires $\mathrm{O}(N)$ operations. Hence, the grid-based filter method is only suitable for case with stationary targets as in~\cite{cliff2015online} where the most expensive computational step, the prediction step, is skipped. Moreover, as shown in Table. \ref{table_MC_planning}, our planning algorithm based on R\'{e}nyi divergence is superior to the Shannon entropy approach in~\cite{cliff2015online} in terms of two important metrics: \textit{i)} accuracy; and \textit{ii)} UAV flight time.  

\begin{table}[!tb] 
	\centering
	\caption{Comparison between our system and \cite{cliff2015online} system.} 
    \vspace{0.2cm}
	\begin{tabular}{L{3cm} |  L{6cm} | L{6cm}} 
		\hline
		& Ours & \cite{cliff2015online}\\ 
		\midrule\midrule
		\textit{Payload} (g) & 260 & 750 \\
		\textit{Total mass (g)} & 1,280 & 2,200 \\
		\textit{Drone type} & Quadcopters (smaller drone)& Octocopters (relatively larger drone) \\[1ex] 
		\textit{Receiver Architecture} & Software defined radio (digital-based, rapidly scan multiple frequencies to support detecting signals from multiple animals) & Analog filtering circuit and a fixed frequency narrowband receiver (analog-based, difficult to re-configure for a new frequency)\\[2ex]
		\textit{Antenna elements} & Compact, lightweight, folded 2-element Yagi antenna (designed for small drone form factor) & Antenna array structure requiring a large spatial separation of two antenna elements and wire ground plane\\[1ex]
		\textit{Measurement model} & Range-only (exploiting the simplicity of a range-only measurement system)& Bearing-only (antenna array, and UAV rotation at grid points with a phase difference measurement system)\\[1ex] 
		\textit{Filtering method} & Particle filter ($\mathrm{O}(N)$ operations per iteration)& Grid-based filter ($\mathrm{O}(N^2)$ operations per iteration) \\
		\textit{Planning algorithm (reward function)} & R\'{e}nyi divergence  &  Shannon entropy \\
		\textit{Targets dynamics} & Multiple mobile targets & A single stationary target \\
        \textit{Nature of targets} & Radio tags carried by humans test subjects & A radio-tagged bird (Manorina Melanocephala) \\
		
		\hline
	\end{tabular}
	\label{table_Field_Experiment_System}
	
\end{table}

The studies in ~\cite{dos2014small} and ~\cite{vonehr2016software} also used an SDR receiver and considered the problem of detecting multiple VHF radio-tag signals using a software defined radio based receiver. We can make the following observations regarding the other SDR based receiver approaches:
\begin{itemize}
\item The team in~\cite{dos2014small} used an SDR payload on a UAV flying a pre-defined flight path to store raw signal detections. This data was post-post processed after the flight to build a signal heat map. The detection range reported in~\cite{dos2014small} is $240$ m, similar to our range of $320$ m.
\item This study in~\cite{vonehr2016software} discussed two software defined radio methods to collect VHF signal measurements: \textit{i)} using the Doppler effect; \textit{ii)} bearing measurements obtained by rotating a drone-mounted Yagi antenna, the so-called Yagi Rotation Methodology. Notably, this measurement approach is like that proposed in \cite{cliff2015online}. Only the Yagi Rotation Methodology was implemented with a reported bearing measurement accuracy of $\pm$30 degrees. More significantly, the detection range reported in ~\cite{vonehr2016software} is up to 1.5 km. This is mainly due to a higher gain antenna (3-element Yagi vs 2-element Yagi of our system) and a more sensitive SDR, the Funcube Dongle Pro+ (FDP+) SDR used in the study. Although the Funcube Dongle Pro+ (FDP+) has a higher receiver sensitivity, it has a limited bandwidth compared the HackRF One SDR device we employed.
\end{itemize}
The mass of the sensor systems was not reported in ~\cite{vonehr2016software}, but Funcube Dongle Pro+ (FDP+) SDR device with a mass of 17~g is significantly more lightweight than the HackRF One we employed with a mass of 100~g. Although detection range cannot be directly compared, we can see that together with a higher gain antenna, the hardware employed in ~\cite{vonehr2016software} achieved a significantly larger signal detection range compared with our study and the the studies in~\cite{dos2014small} and~\cite{cliff2015online}.

\subsection{Lessons Learned}\label{sec:lessons}
In this section, we share our observations and discuss lessons learned during our extensive set of field experiments. We also share with the research community guidelines for establishing a framework for UAV operations and related research.

We realize that the field trials are difficult for any robotics system, especially for aerial platforms where several strict regulations govern their operation. These regulations can depend on jurisdictions under which the flight operations are conducted. Typically, regulations imposed can be different depending on the purpose of the flight such as commercial or recreational and the weight class of the UAV. Currently, there is a lack of harmonization in these regulations. For instance, the requirement for a remote pilot license (RePL) applies to countries such as Australia, Germany, and India only for UAVs over 2~kg ~\cite{casaac101,GermanyBMVI2017,IndiaDGCA2018}. In contrast, New Zealand and Finland only require a license for UAVs over 25~kg~\cite{NewZealandCAA2015,FinnishTrafi2016}. Therefore the research team must \textit{first} familiarize themselves with existing regulations governing the operation of UAVs. Second, the research team needs to negotiate with the insuring body under which they operate to allow the conduct of drone-based flights as this should not be assumed. Insurance agencies can place further restrictions upon the possible field trials that can be conducted due to legal and risk issues. Dealing with these critical issues first will allow getting a framework under which to operate UAV related research such as our work in this article. At the time of doing this research, such a framework was pioneered at our University. This included the creation of a Chief Remote Pilot position and a Maintenance Controller Position. Subsequently, applying to CASA (Civil and Aviation Services, Australia) to obtain a Remotely Piloted Aircraft Operator's Certificate (ReOC) to conduct UAV missions. The Chief Remote Pilot registered with CASA then has the authority to evaluate, manage and approve all UAV flights conducted by University staff and students.

We observed, in both field experiments and simulations, that flying the robot platform higher allows obtaining a better signal compared to ground-based systems. This is because the signal propagating to the UAV system entering an open airspace will be less attenuated than a signal propagating to a ground-based antenna and receiver system. This is since a signal propagating to a ground-based receiver will be more attenuated from potentially multiple radio wave scatters, reflectors, absorbers such as shrubs and grass in the intervening paths. Therefore, flying the robot at a higher altitude can increase the detection range. Notably, in practice, this height advantage is sometimes obtained by using lightweight aircraft and this is an expensive proposition.

The detection range of our current system is not comparable to handheld systems. However, we can see that to develop a mature tool that can function independently and survey a large area of land, we need a longer signal detection range. One simple approach to increase the range is to employ a preamplifier stage for the SDR we have used. An alternative approach is to consider an SDR device with greater sensitivity in the VHF band. For example, an earlier SDR based design ~\cite{vonehr2016software} has achieved a 1.5 km detection range. Although we could not have benefited from such a long range given the limited University allocated space for testing, the study in ~\cite{vonehr2016software} shows that a different SDR device based receiver can offer much longer detection range. Most notably, the SDR used in ~\cite{vonehr2016software} with a mass of only 17 g can be used to replace the SDR of mass 100~g we have employed to realize a further reduction in the mass of the sensor system.

The current flight time for 3DR IRIS+ quad-copter carrying our sensor system is only around 10 minutes while the detection range of the type of VHF collar we have used is around 320 m. Thus, surveying a larger area in the order of several hundred hectares is not yet feasible for our battery-equipped UAV. However, assuming we employ the SDR receiver used in ~\cite{vonehr2016software}, we can achieve a reported detection range of 1.5~km. Consequently, we can see that such a detection range can achieve a survey area defined by a radius of 1.5 km to yield an area of over 700 hectares. Alternatively, if we assume that the survey area scales with the square of the detection range, we can see that an area of 225 hectares can potentially be surveyed.

Further, we observe that flying the UAV close to highly mobile targets helps to reduce localization uncertainty. We can clearly observe this in our path planning results in Fig.\ref{fig_Field_Trial_5tags}a where target $1$ was running back and forth compared with the UAV trajectory for Fig.\ref{fig_Field_Trial_5tags}b where target $1$ was less mobile. However, a close approach by a UAV may disturb the wildlife of interest ~\cite{hodgson2016best,mulero2017unmanned} and can be potentially counterproductive when attempting to obtain accurate spatial and temporal information of threatened species. Wildlife reactions to a UAV differ among different species. For example, terrestrial mammals are less reactive to a UAV than birds~\cite{mulero2017unmanned}. Therefore, the potential for disturbance as well as operating parameters of a UAV close to wildlife is more likely to be dependent on the species of interest. We hope to be able to address questions around appropriate operating parameters for drones in our future work. Nevertheless, we should consider maintaining a safe distance from wildlife. A practical solution can be found by flying at the highest altitude possible~\cite{mulero2017unmanned}. A second approach is to use a receiver with a higher sensitivity, such as the hardware used in~\mbox{\cite{mulero2017unmanned}}, to increase the signal detection range. A third approach can be to reformulate the trajectory planning algorithm using the void probability functional proposed in~\cite{beard2017void}. Such a planning method can alter the control decisions of the path planning algorithm to avoid approaching wildlife and always maintain a safe distance.

\subsection{Limitations and Future Work}\label{sec:limiataions}
While we have demonstrated a successful system, our approach is not without limitations.

Although we formulated a three-dimensional (3D) tracking problem---see equation~\eqref{eq_dynamic_model}---our implementation assumed a fixed UAV altitude during the field trials. Therefore the implemented algorithm solved a two-dimensional (2D) tracking problem, that is ideally suitable for tracking and locating endangered species in largely flat terrains and grasslands. 
Consequently, the current approach is not suitable for tracking wildlife in hills or mountainous areas. 

Notably, implementing a 3D tracking algorithm is straightforward given our formulation is already in 3D. Instead of assuming the target's height is fixed, we need to incorporate an additional unknown variable $p^t_z$ for the target height in the target state space described by $\mathbf{x}$ in Section~\ref{sec_trackign_localizing} and estimate the value of this unknown variable together with the 2D coordinate variables, $p^t_x$ and $p^t_y$. We have conducted simulations to evaluate a 3D formulation where the target height is unknown with an initial uncertainty ranging between $\pm10$~m and the UAV altitude is assumed to be known exactly. The simulation results confirm that our tracking and planning algorithm is still able to track and localize multiple radio-tagged targets with unknown heights. However, the practical challenge is that we need to obtain accurate UAV altitude measurements to implement a robust 3D tracking formulation. Commercial off the shelf UAVs such as the 3DR IRIS+ that we used for building our autonomous system employs a barometer to determine height. We observed in flight tests that the height measurement is unreliable, fluctuates over time and often depends on weather conditions; as also observed in ~\cite{szafranski2013altitude,liu2014auto}. Thus, we leave it for future work to address the problem of accurately estimating the altitude of a UAV. Two approaches that can be considered include: \textit{i)} filtering the barometer sensor data using, for example, a Kalman filter~\cite{liu2014auto}; and \textit{ii)} the use of a LiDAR sensor or a radar-based sensor for more accurate height above ground estimations ~\cite{schartel2018radar}. Alternatively, employing the existing implementation on all topographical conditions require a UAV capability to maintain a fixed relative altitude above ground.

While the software defined radio device may be replaced to achieve a greater detection range, as we discussed in Section~\ref{sec:lessons},  future work should focus on the development of new antenna designs. We designed, simulated and built a compact, folded two element Yagi antenna. Further research efforts to investigate antenna design techniques can lead to lightweight higher gain antennas to increase the detection range and survey area.

The range of the $2.4$ GHz wireless link we employed for communicating between the Ground Control System and the UAV has limited outdoor range---see Figure~\ref{fig_Combine_PC_UAV_Communication_And_Antenna_Design}. Although, this is not a problem given the limited test site available for our work, building a practical tool requires addressing this potential problem. Thus, future work should piggyback data on the telemetry channel using the long-range $915$ MHz radio channel~\cite{vonehr2016software}. Alternatively, the Ground Control System can be removed from the loop by embedding all of the tracking and planning algorithm on the UAV itself to increase the system reliability and search area by eliminating the transmission power consumed by the additional 2.4~GHz radio channel.

Our problem formulation assumes that at least one target is visible or the UAV's initial heading can be in the general direction of the targets. This approach is similar to that followed in \cite{cliff2015online}. In future work, planning formulation should consider both exploration and tracking to deal with events where there are no detectable radio signals ~\cite{charrow2015active}.  

\section{Conclusions} \label{sec:conclusion}
We have developed and demonstrated an autonomous aerial vehicle system for tracking and localizing VHF radio-tagged animals using noisy RSSI based measurements and considered the mobility of targets during their discovery in the field. The joint particle filter and POMDP with R\'{e}nyi divergence based reward function provided an accurate method to explore, track and locate multiple animal collars while considering the resource constraints of the underlying UAV platform. In addition, we have realized a lightweight sensor system to minimize the payload on a UAV and achieve longer flight times.

We have demonstrated the robustness and scalability of the system in field experiments with five VHF radio collar tags under various motion dynamics. We conducted 20 autonomous flights and over 10 manual flights for sensor system evaluations. Our future goal is to evaluate our aerial robot system in field trials with different species of animals. We are in the process of obtaining ethics clearance for our first trial with the engendered Southern Hairy Nose wombats in South Australia.

\subsubsection*{Acknowledgments}
This work was jointly supported by the Western  Australia  Parks and  Wildlife (WA Parks),  the  Australian  Research  Council  (LP160101177),  The Shultz Foundation, the Defense Science and Technology Group (DSTG), and the University of Adelaide's Unmanned Research Aircraft Facility. We would like to thank the support and guidance provided by Mr. Adam Kilpatrick, Chief Remote Pilot and Maintenance Controller at the University of Adelaide, for making the field trials possible and and Remote Pilot, Mr. Fei Chen, Auto-ID Lab, The University of Adelaide for support provided in conducting all of the field experiments in the study. We would like to thank conservation biologist Dr. David Taggart for helping us source the additional VHF collar radio tags as well as the support and guidance provided by Mr Keith Morris, Department of Biodiversity, Conservation and Attractions, Western Australia.
	

\end{document}